# Chemical tuning of spin clock transitions in molecular monomers based on nuclear spin-free Ni(II)

Marcos Rubín-Osanz,[a] François Lambert,[b] Feng Shao,[b,†] Eric Rivière,[b] Régis Guillot,[b] Nicolas Suaud,[c] Nathalie Guihéry,[c] David Zueco,[a] Anne Laure Barra,[d] Talal Mallah,*[b] and Fernando Luis*[a]

[a]Instituto de Nanociencia y Materiales de Aragón, CSIC- Universidad de Zaragoza, 50009 Zaragoza, Spain. E-mail: fluis@unizar.es
[b]Institut de Chimie Moléculaire et des Matériaux d'Orsay, CNRS, Université Paris-Saclay, 91405 Orsay Cedex, France. E-mail: talal.mallah@universite-paris-saclay.fr
[c]Laboratoire de Chimie et Physique Quantiques, Université Paul Sabatier, 31062 Toulouse Cedex 4, France.
[d]Laboratoire National des Champs Magnétiques Intenses, CNRS-Univ. Grenoble-Alpes, 38000 Grenoble, France.
[†]Present address Key Laboratory of Marine Chemistry Theory and Technology, Ministry of Education, College of Chemistry and Chemical Engineering, Ocean University of China, Qingdao 266100, China.
*Email: fluis@unizar.es; talal.mallah@universite-paris-saclay.fr

We report the existence of a sizeable quantum tunnelling splitting between the two lowest electronic spin levels of mononuclear Ni complexes. The level anti-crossing, or magnetic "clock transition", associated with this gap has been directly monitored by heat capacity experiments. The comparison of these results with those obtained for a Co derivative, for which tunnelling is forbidden by symmetry, shows that the clock transition leads to an effective suppression of intermolecular spin-spin interactions. In addition, we show that the quantum tunnelling splitting admits a chemical tuning via the modification of the ligand shell that determines the crystal field and the magnetic anisotropy. These properties are crucial to realize model spin qubits that combine the necessary resilience against decoherence, a proper interfacing with other qubits and with the control circuitry and the ability to initialize them by cooling.

# Introduction

Magnetic molecules are attractive candidates to encode spin qubits and qudits.[1-6] Each molecule consists of a core of one or several magnetic ions, surrounded by non-magnetic ligands. The interaction of these ions with their local coordination sphere determines the magnetic energy levels and states of the molecule. Tuning the molecular structure and the local coordination of the magnetic ions introduces the possibility of designing systems adapted to specific applications,[7-15] and constitutes one of the most appealing characteristic traits of a chemically driven approach.

In principle, any magnetic molecule with two well resolved low-lying energy levels can encode a qubit, provided that one can induce transitions between them by means of external stimuli, often resonant electromagnetic pulses. A further condition is that the qubit states are sufficiently robust against any other interaction that can perturb their coherent evolution. In magnetic molecules and at temperatures that are sufficiently low to "freeze" spin-lattice relaxation, decoherence is mainly associated with magnetic noise, arising from either nuclear spins in the ligands or, in not very diluted samples, from the coupling (e.g. dipolar) to other molecules.[7,8,16-22]

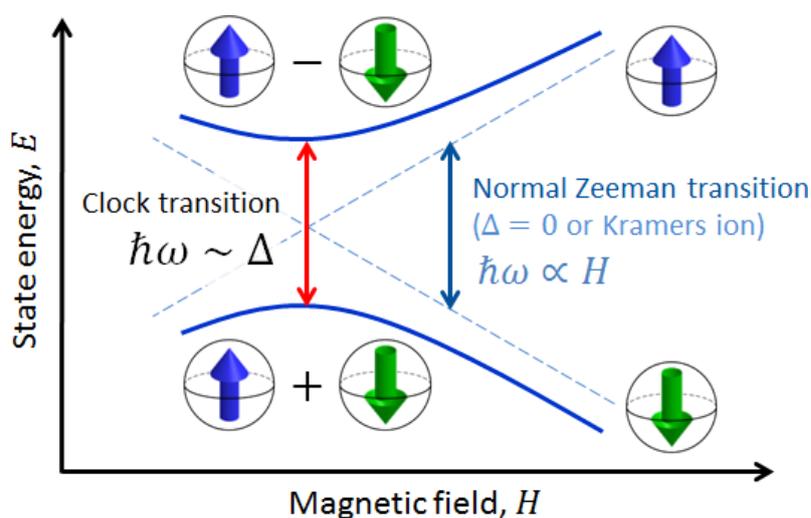

**Figure 1.** Scheme of a spin clock transition. The qubit states are encoded in two superposition spin states that arise at the level anti-crossing. Due to this superposition, the qubit states are protected from magnetic field fluctuations.

Finally, scaling up beyond a qubit, which is obviously necessary for developing any application of real value, implies connecting several qubits. Therefore, a crucial challenge is how to combine a good isolation from noise with the ability to externally control the molecular spin states and their mutual interactions.

A solution to this problem, borrowed from atomic Physics, is to encode the qubit in superposition states that form at level anti-crossings, also known as clock transitions (Fig. 1).[23-27] These level anti-crossings arise from quantum tunnelling between opposite orientations of non-Kramers (i.e. integer) spins.[28] Clock transitions have a remarkable stability against magnetic field fluctuations, as dipolar decoherence vanishes at first order. Examples found among molecular materials[24-26] show that spin coherence times increase sharply near each clock transition. However, the presence of sizeable hyperfine interactions turns them into non-ideal qubit candidates. First, getting to the clock transition requires the application of an external magnetic field and second and, probably more important, the two levels involved in the clock transition might not include the actual ground spin state, thus hindering a simple qubit initialization by cooling.

In this work, we study spin clock transitions in a closer to ideal situation, afforded by molecular [M(Me$_6$tren)Cl](ClO$_4$) monomers, with M = Ni(II) (**1**) and Co(II) (**2**).[29,30] With its $S = 1$ ground state and strongly uniaxial anisotropy, (**1**) provides a realization of the simplest non-Kramer system, with just three magnetic levels ($m_S = 0$ and $\pm 1$). In addition, the only isotope with a nonzero nuclear spin ($^{61}$Ni, $I = 3/2$) has a close to negligible natural abundance of 1.14%. We probe the magnetic level structure by means of heat capacity experiments. The population of quantized levels gives rise to a characteristic Schottky anomaly in the specific heat.[31] This technique is therefore a relatively simple, yet direct, tool to detect small energy splittings, in the range of 0.1 K to 8 K.[32,33] In addition, it allows following its dependence on magnetic field, which is one of the characteristic traits of a clock transition. In complex **2**, isostructural to **1**, Ni(II) is replaced by Co(II), a Kramers ion ($S = 3/2$) for which quantum tunnelling is forbidden by time-reversal symmetry.[28] This allows us to compare both situations and experimentally probe how clock transitions affect spin-spin intermolecular interactions in the same crystal structure.

Finally, we have also studied [Ni(2-Imdipa)(NCS)](NCS) (**4**), a new complex with octahedral geometry close to the strict $O_h$ symmetry that has a correspondingly weaker magnetic anisotropy. This system helps us to show that the quantum tunnelling gap $\Delta$ can be chemically tuned to make it compatible with the microwave circuits that are regarded as promising quantum computation platforms.[34-36]

**Experimental methods**

**Synthesis and structures**

[M(Me$_6$tren)Cl](ClO$_4$) with M = Ni (**1**), Co (**2**) and Zn (**3**); [Ni(2-Imdipa)(NCS)](NCS) (**4**). Molecules **1**, **2** and **3** are isostructural and have been previously reported and studied,[29,30] while **4** is a new complex designed and reported here as a demonstration of chemical tuning of the tunnelling gap of **1**. The details of its synthesis and crystal structure are given in the accompanying Supplementary Information (SI). Solid solutions of, nominally, 10% **1** and 10% **2** into the diamagnetic **3** have also been prepared. They are denoted by **1$_{d9.5\%}$** and **2$_{d11\%}$**, respectively. The actual concentrations were determined by comparing magnetization isotherms measured on the pure and diluted samples under identical conditions (Fig. S1 of the SI).

**Heat capacity experiments**

The specific heat of all complexes was measured, between $T$ = 0.35 K and 20 K (100 K for **3**), with a commercial physical property measurement system (PPMS-Quantum Design) that makes use of a relaxation method.[32,33] Powder samples of **1**, **2** and **3** were mixed with apiezon N grease to improve the thermal contact with the calorimeter. The powder samples were pressed into a small pellet, in the form of a thin disk which was then placed in contact with the calorimeter platform. We find that the crystallites that form the powder have a geometrical tendency to pack in some orientations under pressure. The crystal axes are therefore not randomly oriented but show some preferential orientation in these samples (see Fig. S2 for an estimate). In the case of complex **4,** experiments were performed on single crystals placed onto a thin layer of apiezon N grease.

**Magnetic characterization**

Magnetization data of 4 were measured, in sintered powder form as well as on single crystals, with a commercial Magnetic Properties Measurement System (MPMS-Quantum Design), a magnetometer based on a SQUID sensor, between 1.8 K and 300 K, magnetic fields up to 5 T and for different crystal orientations. The magnetic ac susceptibility of 1 (in powder form) and 4 (on a single crystal and on a powdered sample obtained by crushing the same crystal) was measured using the AC measuring options of the same SQUID magnetometer (for frequencies between 1 Hz and 1.4 kHz) and of the PPMS system (for frequencies between 100 Hz and 10 kHz).

**EPR experiments**

High-Frequency EPR spectra of a single crystal of **1** were recorded on a multifrequency spectrometer operating in a double-pass configuration. A 110 GHz frequency source (Virginia Diodes Inc.) was used. The power of the frequency source was varied, either with an external attenuator or with internal one. The emitted power was measured independently by a powermeter. The exciting light was propagated with a Quasi-Optical bridge (Thomas Keating) outside the cryostat, working in induction mode which provides an efficient isolation of the detector, and with a corrugated waveguide inside it. The detection was carried out with a hot electron InSb bolometer (QMC Instruments). The main magnetic field was supplied by a 16 T superconducting magnet associated with a variable temperature insert (Cryogenic).

*Ab initio* **calculations**

*Ab initio* calculations were performed using the ORCA 4.0.1.2 quantum chemistry package.[37] The *D* and *E* parameters of the new complex **4** were evaluated following the procedure already developed by some of us.[38] State Average CASSCF (Complete Active

Space Self Consistent Field) calculations were first performed to account for non-dynamic correlations; then, dynamical correlations were accounted for using the NEVPT2 method in its strongly contracted scheme.[39-41] Finally, Spin-Orbit (SO) couplings were calculated using the Spin Orbit State Interaction (SI-SI) method implemented in ORCA.[42] The Complete Active Space (CAS) is composed of the five mainly-3d orbitals of the Ni ion and the 8 associated electrons, i.e. CAS(8,5). The averaged CASSCF molecular orbitals optimization was done over the 10 triplet and 15 singlet spin states generated by the CAS(8,5). The SO coupling was considered between all the $m_S$ components of these states, the spin-free energy (diagonal elements of the SO matrix) being evaluated at the NEVPT2 level.

The DKH-def2-QZVPP basis sets were used for Ni (14s10p5d4f2g), for S atoms (11s7p4d2f1g), for N atoms (8s4p3d2f1g) belonging to the coordination sphere of Ni and for the C atoms (8s4p3d2f1g) of NCS$^-$, while for other N and C atoms the DKH-def2-TZVP (6s3p2d1f) basis was used and a def2-SVP (2s1p) basis for H atoms. For DFT calculations, df2-SVP atomic basis sets were used for Ni (5s3p2d1f), for C and N atoms (3s2p1d), for S (4s3p1d) and for H (2s1p).

## Results and discussion

### Direct detection of spin clock transitions by heat capacity experiments

Figure 2 shows the specific heat $c/R$ of **1** measured at zero magnetic field. Above approximately 8 K, it is dominated by the contribution arising from lattice vibrations. This contribution has been determined independently from the specific heat of the diamagnetic complex **3**, which is also shown in Fig. 2. At low and very low temperatures the specific heat of **1** shows an additional Schottky-like anomaly, with a maximum centred at $T_0$ = 1.75 K.

This anomaly signals the existence of a finite gap in the spectrum of magnetic energy levels of this complex.[32,33] Since 98.86% of the stable Ni isotopes carry no nuclear spin, the anomaly must be associated with thermal excitations of the molecular electronic spins. In order to gain further information on its nature, we have measured a sample (**1**$_{d9.5\%}$) that contains 9.5% of **1** diluted into the diamagnetic complex **3**. Despite the expected decrease in the strength of intermolecular magnetic interactions, these data show the same Schottky anomaly, even shifted to slightly higher temperatures (see also Fig. S3). Therefore, we can safely discard that spin-spin interactions originate the Schottky-like anomaly observed in **1** and conclude that it is due to a zero-field splitting (ZFS) intrinsic to each individual molecule.

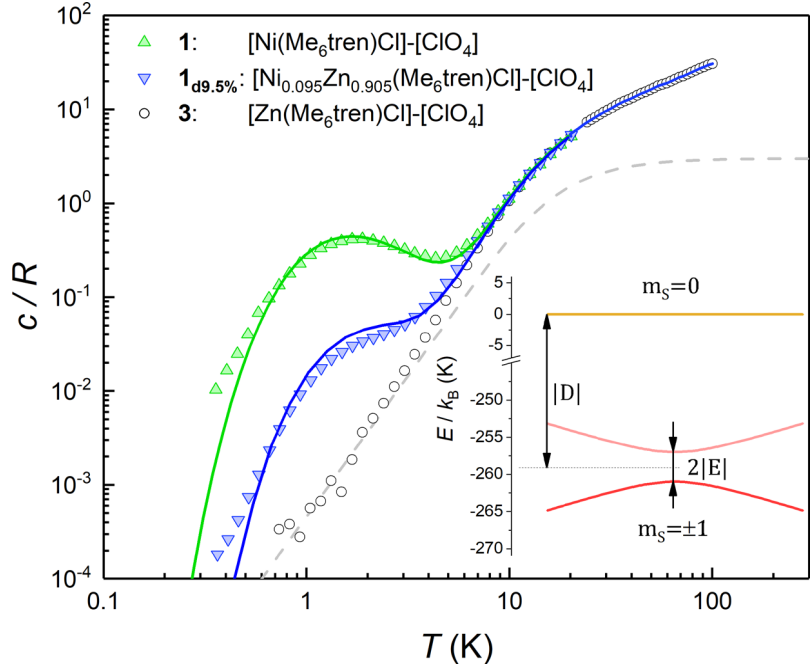

**Figure 2.** Specific heat of complexes **1**, **1**<sub>d9.5%</sub> and the diamagnetic **3** at $\mu_0 H = 0$. The latter gives the lattice contribution to the specific heat. Solid lines show the simulation of the magnetic contribution for a powder sample, plus the experimental lattice contribution from complex **3**. The dashed line shows the contribution of acoustic modes as given by the Debye model with a Debye temperature $\theta_D = 72$ K. *Inset*: Energy levels arising from the anisotropy and Zeeman terms in the effective spin Hamiltonian (1), for both **1** and **1**<sub>d9.5%</sub>. At low temperatures only the subspace spanned by the $m_S = \pm 1$ states is thermally populated and contributes to the heat capacity. The very large $|D|$ diminishes the thermal population of the $m_S = 0$ state and its mixing with the $m_S = \pm 1$ states.

These results can be easily understood as follows. We show in Fig. 2 the specific heat of **1** (solid lines) and the energy levels (inset) calculated by using the spin Hamiltonian reported in [3]:

$$\mathcal{H} = \mu_B \vec{H} \tilde{g} \vec{S} + D S_z^2 + E(S_x^2 - S_y^2) \tag{1}$$

where $\tilde{g}$ is the gyromagnetic tensor that determines the Zeeman interaction with the magnetic field $\vec{H}$, and $D$ and $E$ are the diagonal (or uniaxial) and off-diagonal (rhombic) magnetic anisotropy constants (ZFS parameters), respectively. Because of its very strong uniaxial anisotropy, the magnetic energy level spectrum of **1** shows a highly excited $m_S = 0$ level above the ground $m_S = \pm 1$ doublet. The Schottky anomaly associated with the population of the former level is expected to show up above 40 K, thus it is completely masked by the much higher lattice contribution and, therefore, experimentally undetectable. To all practical purposes, the behaviour of **1** at low temperatures reduces to that of a text book two-level system.[31] The gap observed experimentally must then arise from the weaker off-diagonal anisotropy term $E(S_x^2 - S_y^2)$. This term induces tunnelling between the $m_S = \pm 1$ sub-levels and gives rise to a quantum tunnelling gap $\Delta = 2E$ and to a Schottky-like specific heat contribution, as shown in Fig. 2.

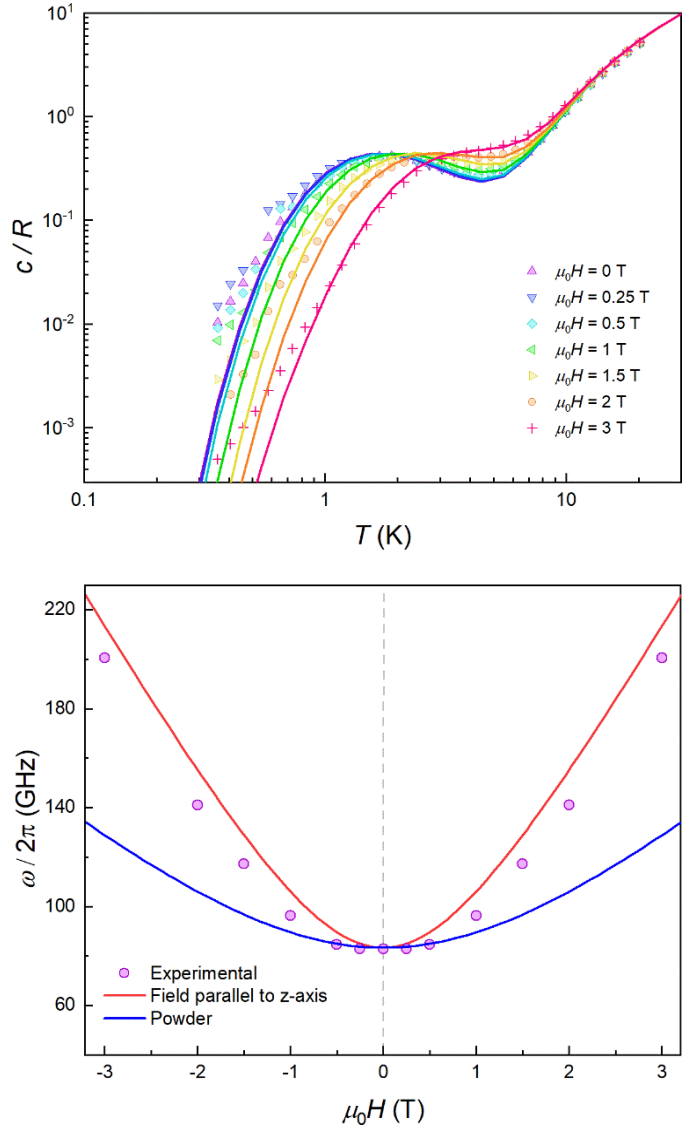

**Figure 3.** *Top*: Specific heat of a powder sample of **1** at $\mu_0 H = 0$ (purple), 0.25 T (blue), 0.5 T (light blue), 1 T (green), 1.5 T (yellow), 2 T (orange) and 3 T (red). Solid lines show the simulation for a powder sample with a preferential orientation along the molecular *z* axis. *Bottom*: Experimental effective energy gap $\hbar\omega = k_B T_0/0.42$, with $T_0$ the field-dependent temperature of the heat capacity maximum. Solid lines show the behaviour expected for a magnetic parallel to the magnetic anisotropy axis *z* (red) and for a randomly oriented sample (blue).

The temperature $T_0$ of the specific heat maximum provides a simple and direct method to determine $\Delta$. At zero field, $T_0$ of a simple two level system is given by[31,33]

$$\frac{\Delta}{2k_B T_0}\tanh\left(\frac{\Delta}{2k_B T_0}\right) = 1 \Rightarrow$$

$$k_B T_0 \simeq 0.42\,\Delta \qquad (2)$$

from which we estimate Δ = 2.9 cm$^{-1}$ (or 83.5 GHz) and $E = \Delta/2 = 1.45$ cm$^{-1}$, which agrees fairly well with the value of 1.6 cm$^{-1}$ estimated from high-frequency EPR experiments.[29] Using this $E$ value, we reproduce very accurately the experimental specific heat data of **1** and **1$_{d9.5\%}$**. The two techniques (EPR and heat capacity) complement each other to provide a full characterization of the magnetic anisotropy.

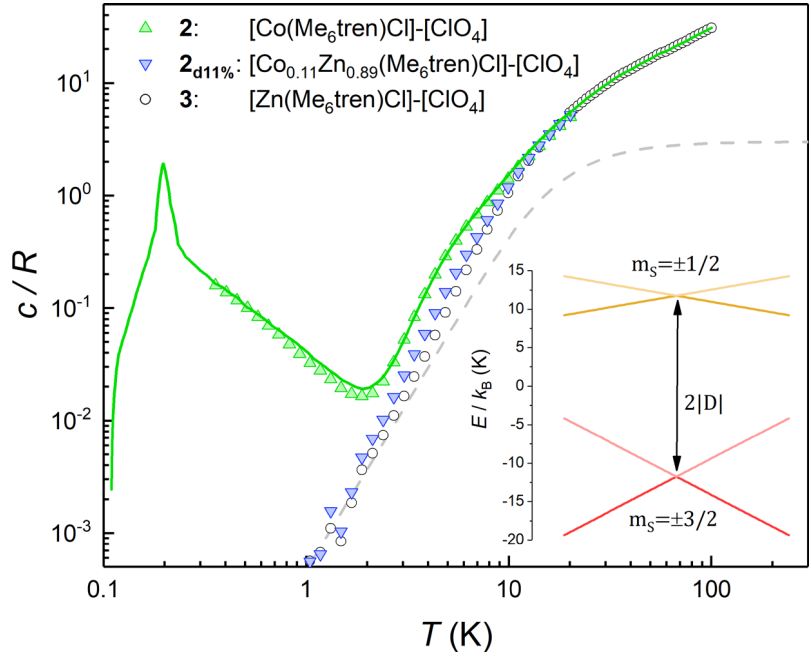

**Figure 4.** Heat capacity of **2**, **2$_{d11\%}$** and the diamagnetic **3** at $\mu_0 H = 0$. The theoretical specific heat (green solid line) includes the contribution arising from spin-spin interactions. The dashed line shows the contribution of acoustic modes as given by the Debye model with a Debye temperature $\theta_D = 72$ K. *Inset*: Energy levels of **2** and **2$_{d11\%}$** calculated with the effective spin Hamiltonian (1) without spin-spin interactions. At low temperatures only the $m_S = \pm 3/2$ doublet contributes to the heat capacity. Above approximately 2 K the population of the $m_S = \pm 1/2$ levels shows up as an extra contribution over the lattice specific heat.

The quantum nature of the observed level splitting is confirmed by its magnetic field dependence. Figure 3 (top) shows specific heat data of **1** measured for $0 \leq \mu_0 H \leq 3$ T. The low-$T$ anomaly hardly changes for magnetic fields below 1 T. A weak dependence on magnetic field is precisely the behaviour expected for the energy gap $\hbar\omega$ near a spin clock transition. Near the anti-crossing, $\hbar\omega$ is approximately given by

$$\hbar\omega = \sqrt{(2g_z\mu_B S H_z)^2 + \Delta^2} \qquad (3)$$

The temperature $T_0$ of the specific heat maximum can then be calculated by simply replacing Δ in Eq. (2) by this field-dependent $\hbar\omega$. It then follows that the magnetic field has little influence on the specific heat provided that $2g_z\mu_B SH_z \ll \Delta$, as observed experimentally. The parabolic dependence of the quantum level splitting predicted by Eq. (3) can be directly monitored by plotting the experimental $T_0$ as a function of $H$. The data nicely follow this dependence, as shown by the bottom panel of Fig. 3. Yet, the experimental shift of the heat capacity anomaly is larger than what one would expect for a randomly oriented sample, suggesting that the polycrystalline sample has a preferred orientation, as expected (see methods). The orientation that best accounts for the heat capacity data is a "mix" between that of a random powder and a fully oriented sample with *z* parallel to the magnetic field (cf Fig. 3 and Fig. S2).

**Comparison with a Kramers spin system: effect of spin clock transitions on spin-spin interactions**

This section describes a realization, within the same molecular system, of a spin with a conventional level crossing (cf Fig. 1), thus providing a nice comparison with the "quantum limit" that complex **1** represents. For this purpose we study the Co(II) containing complex **2**. Whereas Ni(II) has an integer spin $S = 1$, Co(II) possesses an $S = 3/2$ ground state. Therefore, it is a Kramers ion. Quantum tunnelling of the electronic spin is then strictly forbidden on account of time-reversal symmetry.[28] We expect to see profound differences in the physical behaviour of these two derivatives. As it is shown in the following, these differences manifest themselves neatly in the heat capacity.

Figure 4 shows the specific heat of **2** measured at zero field. The magnetic anisotropy of **2** is much weaker than that of **1**.[30] For this reason, in this case the anomaly associated with the ZFS between the $m_S = \pm 1/2$ and $m_S = \pm 3/2$ level doublets (see the inset of Fig. 4) shows up at lower temperatures and can be observed as an extra contribution to $c/R$ over the lattice contribution. The experimental results agree with simulations performed using the spin Hamiltonian in Eq. (1) with $D = -8.31$ cm$^{-1}$ and $E \approx 0$, which result in an overall ZFS of $2D$. These values are in almost perfect agreement with $D = -8.12$ cm$^{-1}$ and $E = 0$ determined from high-field EPR.[30]

The additional contribution to $c/R$ that is observed below 1 K must therefore arise from the splitting of the ground $m_S = \pm 3/2$ doublet. This contribution can, in principle, be associated with either hyperfine interactions (Co only stable isotope has $I = 7/2$ nuclear spin) and/or with spin-spin couplings between different molecules in the crystal. However, as it is shown in Fig. S4, the former are expected to be too weak to account for the specific heat measured below 1 K. In order to confirm this, we have measured the heat capacity of a magnetically diluted sample **2$_{d11\%}$** of **2** in the diamagnetic derivative **3**. The low-*T* contribution observed for the pure compound **2** is greatly suppressed in **2$_{d11\%}$**, i.e. not just rescaled by concentration, and virtually disappears. This result confirms that isolated molecules of **2** have no electronic gap, as expected, and suggests that the low-*T* specific heat reflects mainly the effects of intermolecular interactions.

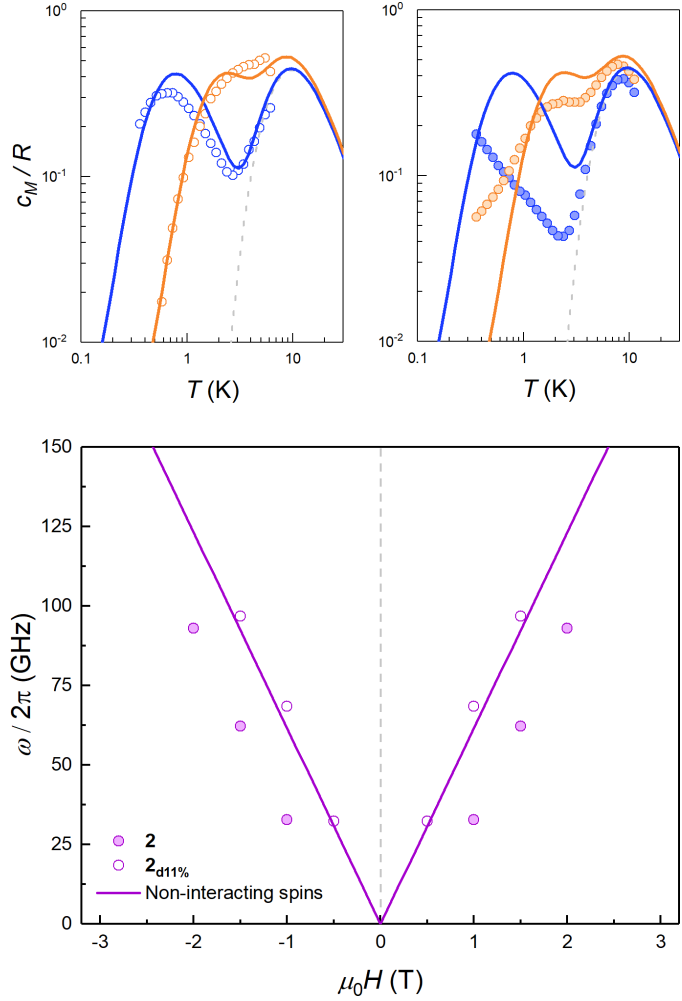

**Figure 5.** *Top*: Magnetic specific heat $c_m$ of $2_{d11\%}$ (open dots, left panel) and of **2** (solid dots, right panel) at selected fields $\mu_0 H = 0.5$ T (blue) and $1.5$ T (orange) obtained after subtraction of the lattice contribution estimated for the diamagnetic compound **3**. In both cases, $c_m$ is normalised per mol of magnetic molecules. The dashed line shows the zero field splitting contribution (associated with the $2|D|$ gap in the inset of Fig. 4), while solid lines show the simulation of the expected heat capacity for non-interacting molecular spins (ZFS + Zeeman interaction). *Bottom*: Experimental energy gap $\hbar\omega = k_B T_0/0.42$, with $T_0$ the field-dependent temperature of the heat capacity maximum. The solid lines show the Zeeman level crossing expected for non-interacting molecules with the magnetic field applied along the z molecular axis.

For a more quantitative description, the effects of intermolecular spin-spin interactions have been calculated by performing Monte Carlo simulations on the crystal lattice. We have used the following simple model:

$$\mathcal{H}_{int} = -\frac{1}{2} J \sum_i \sum_{j \in Z(i)} S_{i,z} S_{j,z} \qquad (4)$$

where index i goes through the whole lattice and index j labels the $Z(i)$ nearest-neighbours of site i, which couple via a constant interaction constant $J$ (see Fig. S5 for further information on the lattice topology). As a further simplification, the spins are restricted to fluctuate only between the two states of the lowest energy doublet $m_S = \pm 3/2$. This assumption is based on the fact that the thermal populations of the excited $m_S = \pm 1/2$ states become negligible in the temperature range below 2 K, to which the simulations apply. As shown by the lines in Fig. 4, we find a good agreement with the experimental data of **2** for $J = -0.035$ cm$^{-1}$. In particular, the model predicts that **2** should undergo a phase transition to an antiferromagnetic phase at about $T_N = 0.22$ K.

The sign of $J$ can be further refined by looking at the field dependence of the specific heat. Data measured on pure and magnetically diluted samples of **2** are shown in Fig. 5. The Zeeman interaction splits the $m_S = \pm 3/2$ levels and competes with the spin-spin interactions. The specific heat then shows a rounded maximum, again characteristic of a two-level system, which progressively shifts towards higher temperatures as $H$ increases. Although, with the proper scaling, data measured on the pure **2** and the magnetically diluted **2$_{d11\%}$** samples tend to approach each other as $H$ increases, the Zeeman splitting, as measured from the maximum temperature, remains always larger for the latter. In addition, the magnetic specific heat of **2$_{d11\%}$** is closer, at any $H$, to what is expected for non-interacting molecules. These results confirm that spin-spin interactions have a predominantly antiferromagnetic character in this lattice, as magnetization hysteresis measured at very low temperatures already suggest.[30]

The results described in this and the previous section allow gaining a deeper insight on how a sizeable quantum tunnelling gap affects spin-spin interactions. Complex **2** illustrates the conventional, or classical, behaviour (cf Fig. 4): interactions between spins in an ordered lattice break the symmetry between spin-up and spin-down states and lead to a transition towards a long-range ordered phase at sufficiently low temperatures. By contrast, and in spite of having the same crystal lattice, the specific heat of complex **1** (cf Fig. 2) does not show, below 0.5 K, any signature of spin-spin interactions. The close to perfect scaling with spin concentration, and the dependence on magnetic field (cf Fig. 3) suggest, instead, that these molecules behave as quasi-isolated.

In order to illustrate the essential ingredients that govern this "quantum decoupling", we have numerically diagonalized the spin Hamiltonian $\mathcal{H} + \mathcal{H}_{int}$, where $\mathcal{H}$ and $\mathcal{H}_{int}$ are given by Eqs. (1) and (4), respectively, for an affordable lattice comprising just one $S = 1$ spin and its six nearest neighbours located within the same crystallographic plane. Details of these "toy quantum model" calculations are given in Fig. S6. The results reflect how quantum fluctuations influence the ground state of **1**. At zero field, the wave function becomes a symmetric superposition of spin-up and spin-down states whenever the quantum tunnelling gap $\Delta$ exceeds the characteristic energy scale $\varepsilon \approx Z|J|S^2/2$ of spin-spin interactions. If we estimate $\varepsilon$ of **1** by taking the same $J$ that was determined for complex **2** and $S = 1$, it follows that $\Delta \approx 10\,\varepsilon$ for this system. The condensation into the ground singlet state then largely suppresses intermolecular magnetic interactions.

**Chemical tuning of spin clock transitions**

*Synthetic strategy.* Qubits encoded in spin clock transitions are more stable, but for the very same reason they are also more difficult to tune: from Eq. (3), the effect that the magnetic field has on the qubit frequency ω decreases with increasing Δ. Besides, if quantum operations are to be induced by resonant pulses applied with commercial EPR cavities or on-chip superconducting resonators, one gets limited to specific frequency ranges. In the former case, the most widely used EPR spectrometers operate at 9-10 GHz (X-band) or 35 GHz (Q band).[8-16,18-25] In the case of superconducting circuits, the frequencies range mainly from 1 to 10 GHz.[34-36] We have shown above that **1** possesses a quantum tunnelling gap $\Delta/h \approx$ 83 GHz. It would therefore be highly desirable to find a chemical strategy enabling to develop related molecular systems with smaller tunnelling gaps.

For **1**, the tunnel splitting is due to the Jahn-Teller effect that lifts the orbital degeneracy of the Ni(II) electronic state. This effect distorts the trigonal plane of the complex and gives rise to the appearance of a large rhombic parameter (*E* = 40 GHz), as demonstrated above.

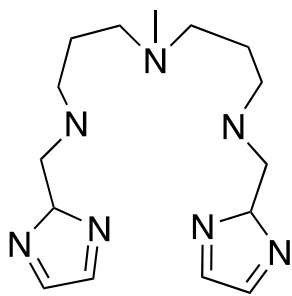

**Figure 6.** Schematic view of the 2-Imdipa pentadentate ligand.

A relatively straightforward strategy to drastically reduce the quantum tunnelling splitting (2*E*) is to globally reduce the axial ZFS parameter *D*, so that the |*E*| parameter will be restricted to an upper limit of (1/3)|*D*|, assuming of course that the two molecules at hand have a similar rhombicity |*E/D*|.

An efficient approach to achieve small |*D*| values, in comparison to that of complex **1**, is to consider a Ni(II) complex with an octahedral geometry, as close as possible to $O_h$ symmetry but not strictly $O_h$. Indeed, for $O_h$ symmetry, the three components of the orbital moment ($L_x$, $L_y$ and $L_z$) belong to the same irreducible representation $T_{1g}$ of $O_h$. The effect of SOC will then be the same in the three axes and the three-fold degeneracy of the ground state ($^3A_{2g}$) cannot be lifted (the three $m_S$ sub-levels 0 and ±1 would have the same energy). However, if the complex is slightly distorted the effect of SOC will be slightly different along each direction of space, resulting in a very low value of |*D*|.[28,43]

With this idea in mind, we prepared several octahedral Ni(II) complexes using a pentadentate organic ligand (2-Imdipa, Fig. 6 and Figs. S7-S9) and different axial ones (NCS⁻, Cl⁻, NO₃⁻). It turned out that the hexacoordinate complex an octahedral geometry bearing NCS⁻ as axial ligand of formula [Ni(2-Imdipa)(NCS)](NCS) (complex **4**) has all the characteristics required as summarized in Fig. 7, i.e. a similar rhombicity ($|E/D|$) but an axial parameter $D$ that is nearly two orders of magnitude smaller (see below for details).

The preparation and full characterization of the organic ligand and complex **4** are detailed in the SI. The compound crystallizes in the space group P2$_1$/n (group 14, see Table S1). There are two crystallographically independent molecules (Figure S7) that mainly differ by about 1° tilts in their relative orientations, and that have almost the same coordination sphere (Figure S8). We will, therefore, focus on one of them.

The geometry around Ni is a distorted octahedron (Figure S9). The three amine nitrogen atoms (N2, N3 and N4) and N1 belonging to one of the imidazolate groups of the pentadentate ligand 2-Imdipa lie almost in one plane (equatorial). The nitrogen atom of the other imidazolate group (N5) and that of NCS⁻ (N6) occupy the apical positions. The Ni-N bond distances in the equatorial planes are all larger (average value 2.137 Å, see Table S2) than the axial ones (2.060 Å). The axial bond distances are identical within experimental errors while those in the equatorial plane differ by less than 0.087 Å (Table S2).

*Magnetic anisotropy determination.* The magnetic data were first collected using a powder sample. The $\chi T$ product (Figure S10) remains approximately constant between room temperature and 20 K, with a value of 1.19 cm³mol⁻¹K in line with a $S$ = 1 state with a $g$-factor of 2.16. Below $T$ = 20 K, $\chi T$ slightly decreases and reaches a value of 0.99 cm³mol⁻¹K at $T$ = 2 K.

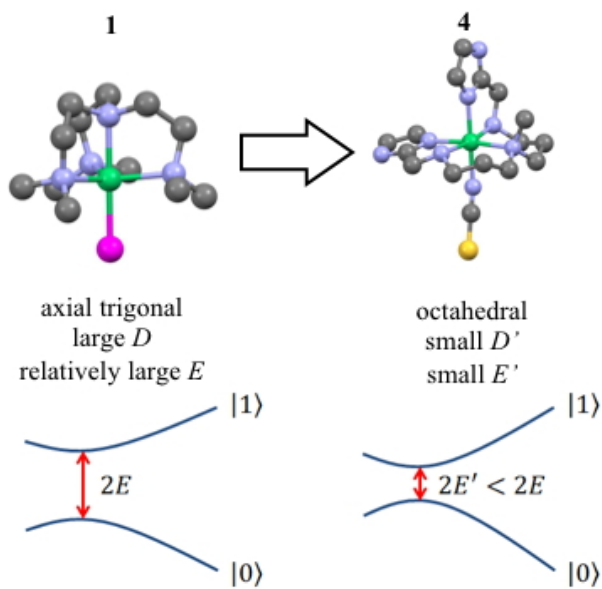

**Figure 7.** Structures of **1** and **4**, with their corresponding quantum tunnelling gaps between the qubit states |0⟩ and |1⟩. The close to O$_h$ symmetry of **4** results in smaller absolute values of the distortion parameters $E$ and $D$, so the gap is chemically tuned towards lower frequencies.

Because the structure does not show any intermolecular path for strong antiferromagnetic couplings, this small decrease can tentatively be attributed to the ZFS of the $S = 1$ state. Magnetization isotherms were measured at $T = 2$, 4 and 6 K.

When plotted against $\mu_0 H/T$ (Figure S11), these magnetization curves are not superimposable, which again confirms the presence of a net ZFS within the $S = 1$ ground state. Fixing the isotropic $g$ factor to 2.16, as derived from $\chi T$ data, it is possible to achieve very high quality fits of the data for negative and positive $D$ values and a TIP equal to $10^{-4}$, as expected for Ni(II) octahedral complexes. The fit parameters are: $D = +2.11$ cm$^{-1}$ and $|E| = 0.09$ cm$^{-1}$ ($|E/D| = 0.04$) when starting the fit with a positive $D$-value and $D = -2.96$ cm$^{-1}$ and $|E| = 0.06$ cm$^{-1}$ ($|E/D| = 0.02$) when starting from a negative $D$ value. These data show that, as expected for a Ni(II) complex with an octahedral geometry, the axial ZFS parameter ($D$) is very small and the rhombic parameter ($E$) that defines the tunnel splitting ($2E$) is also very small. However, magnetization data on a powder sample cannot discriminate between positive $D$ ($m_S = 0$ ground level) or negative $D$ ($m_S = \pm 1$ ground levels).

In order to further elucidate the magnetic anisotropy of complex **4**, we carried out heat capacity and magnetization studies on a single crystal that has the form of a hexagonal prism (Figure S12). The specific heat measured for $\vec{H}$ along the perpendicular to the hexagonal faces, within the *ac* plane, is shown in Fig. 8. This quantity reflects the thermal population of the spin levels, thus it critically depends on the sign of $D$. As Fig. S12 shows, the results are incompatible with a positive $D$, and agree well with a uniaxial anisotropy ($D < 0$). The fits of these data and of those obtained by applying the magnetic field along *b* (see the SI for symmetry considerations used in these fits and Fig. S13 for the results) suggest that the easy magnetization axis *z* lies on the *ac* plane at 52.6° from the perpendicular to the *ab* plane as shown in Fig. 9. Besides, the Schottky anomaly centred near 2 K, which corresponds to the thermal excitation of the $m_S = 0$ state, evidences that **4** has a weaker magnetic anisotropy than **1**, and gives $D \approx -2.71$ cm$^{-1}$.

*Angle-dependent magnetic response.* These conclusions have been checked against magnetization measurements performed for varying crystal orientations. In these experiments, the magnetic field was orthogonal to the rotation axis. Results obtained by rotating a single crystal of **4** around the perpendicular to its *ab* plane are shown in Fig. 9. The magnetic field is then confined within the *ab* plane. We obtain a minimum magnetization when the field points along *b*, which means that this axis is perpendicular to the magnetic anisotropy *z* axis. The maximum magnetization is found for the field pointing along *a*, which forms 37.4° = 90° - 52.6° with the molecular *z* axis. Magnetic susceptibility and magnetization data measured at these two field orientations are given in Figs. S14 and S15, respectively. The results and, in particular, the positions of minima and maxima agree well with those predicted using the anisotropy parameters and the orientations of the anisotropy axes shown in Fig. 9.

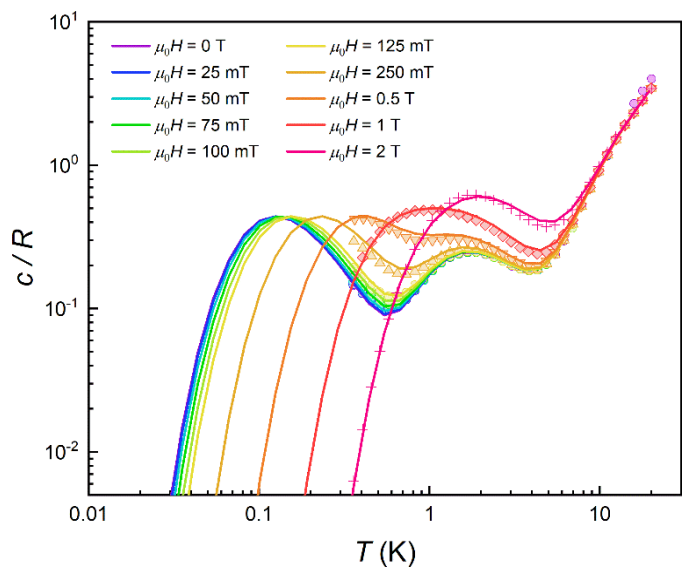

**Figure 8.** Specific heat of a single crystal of **4** at $\mu_0 H = 0$ (purple), 25 mT (blue), 50 mT (light blue), 75 mT (green), 0.1 T (light green), 0.125 T (yellow), 0.25 T (light orange), 0.5 T (orange), 1 T (red) and 2 T (pink). Solid lines show the simulation for a crystal having its magnetic z axis at 52.6° from the magnetic field.

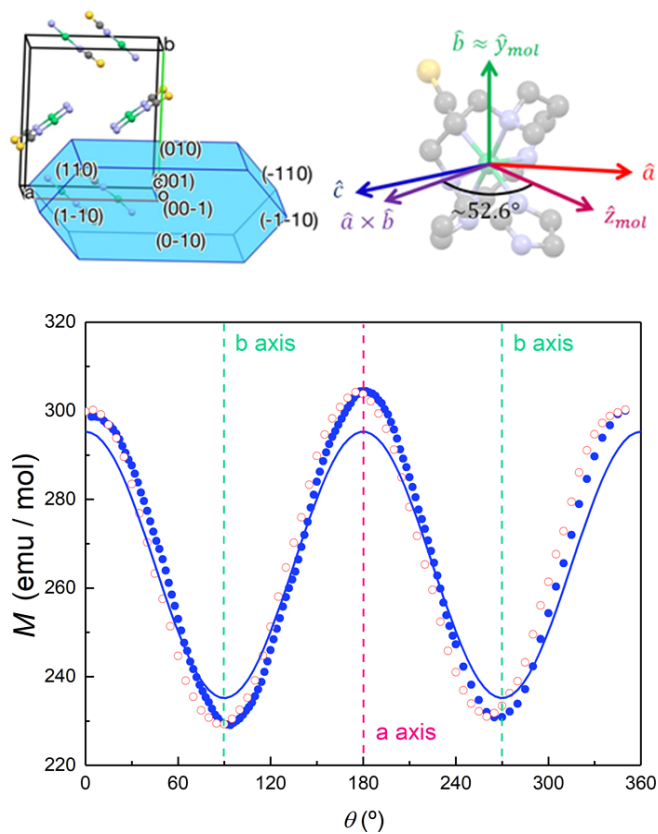

**Figure 9.** *Top left*: relation between the unit cell axes and the crystallisation structure of **4**. *Top right*: relation between the unit cell and magnetic anisotropy axes of **4**. *Bottom*: Magnetization of **4** measured at $\mu_0 H = 0.1$ T and $T = 5$ K as a function of rotation angle around an axis perpendicular to the *ab* unit cell plane. The difference between data measured while increasing (blue solid dots) and decreasing (red open symbols) $\theta$ is due to a mechanical hysteresis of the rotation system and provides a measure of the angular

uncertainties. The crystal is placed so that the field points along the *b* unit cell axis at θ = 90° and θ = 270°. The solid line shows the magnetization simulated, with the anisotropy parameters given in the text, for a crystal (with all four molecule orientations) rotated around the perpendicular to *a* and *b*.

*Ab initio calculations.* Calculations, performed at the NEVPT2 level for the two independent molecules, lead to almost the same ZFS parameters: $D$ = -2.43 and -2.55 cm$^{-1}$ and $E$ = -0.36 and 0.34 cm$^{-1}$ (close to the experimental one (-2.71 cm$^{-1}$ and 0.1 cm$^{-1}$). The analysis of the results, performed in the framework of the magnetic axes frame, shows that the negative value of $D$ is the result of negative and positive contributions of different triplet and singlet excited states that couple to the ground state (see Table S3 for a perturbative evaluation of the contribution of each state). When added, they lead to the small negative value. The orientations of the $D$ tensor principal axes (Figure S16) coincide with those determined experimentally from the magnetization study on a single crystal, described above.

Getting insight of the contribution of each state by analysing the composition of their wave function is possible, as it has recently been done on a Ni(II) complex with an octahedral geometry,[44] but is out of the scope of this paper. These results confirm our reasoning that a Ni(II) hexacoordinate complex with octahedral geometry that slightly deviates from $O_h$ symmetry leads to small $D$ value and consequently small $E$ (as $|E|<|D|/3$). However, because of the very weak deviation from $O_h$ symmetry, it is not possible to predict without ab initio calculations the sign of $D$ that is of major importance for using these systems as qubits.

*Quantum tunnelling gap.* As in the case of complex **1**, the quantum tunnelling splitting can also be "read-out" from the position of the specific heat anomaly observed below 1 K. The data are well accounted for with Δ = 2$E$ ≈ 0.21 cm$^{-1}$ (Δ/$h$ ~ 6 GHz). The $D$ and $E$ experimental values agree rather well with those predicted (see previous section) from *ab-initio* calculations. Repeating this experimental protocol for different $H$ values, it is possible to monitor the magnetic field dependence of the energy gap $\hbar\omega$ of the ground state doublet. As Fig. 10 shows, it follows a quadratic dependence with a reduced zero-field quantum gap Δ as compared to the much more strongly anisotropic complex **1** (compare Fig. 10 to Fig. 3). These results confirm also that the magnetic anisotropy *z* axis makes 52.6° with respect to the magnetic field.

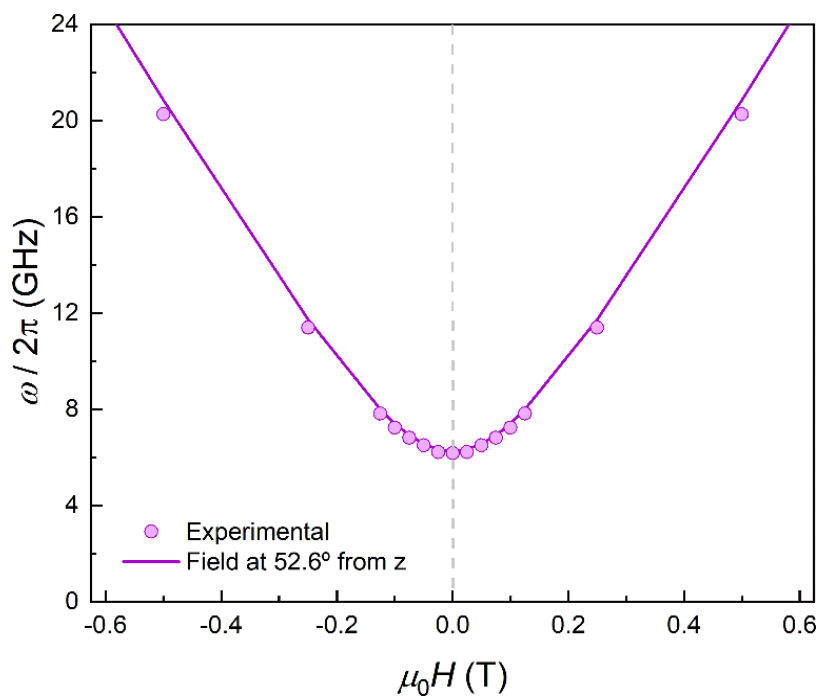

**Figure 10.** Experimental effective energy gap $\hbar\omega = k_B T_0/0.42$, with $T_0$ the field-dependent temperature of the heat capacity maximum. Solid lines show the behaviour expected for a magnetic field applied at 52.6° from the anisotropy axis.

**Spin relaxation**

The data shown in Figs 3 and 10 provide a direct mapping of the anticrossings between the two lowest lying spin levels of **1** and **4**. Besides, we have shown that the zero-field eigenstates become quite insensitive to environmental magnetic fields. These two properties, which are characteristic traits of a spin-clock transition, make these molecules promising spin qubit candidates. However, they might not be enough, if the spin relaxation towards thermal equilibrium, parameterized by the spin-lattice relaxation time $T_1$, becomes also very fast. We have addressed this important question experimentally, by combining frequency-dependent ac susceptibility and EPR measurements.

Illustrative ac susceptibility data are shown in Figs. S17 to S19. They show a dependence on frequency $\omega$ for both **1** and **4**, signalling the presence of relatively slow relaxation processes[45] (note that the characteristic time scale of these measurements is $1/\omega > 16$ μs). In the case of **1**, however, the data do not allow a quantitative determination of $T_1$ because the maximum of the imaginary susceptibility component $\chi''$ (that approximately corresponds to $1/\omega = T_1$) appears to be above our highest attainable frequency. A further limitation arises from the fact that superposition spin states (Fig. 1) have a zero average magnetic moment. As a result, phonon-induced transitions between different eigenstates do not lead to any change in the susceptibility (see Fig. S20).[46] Near the clock transition (< 0.1 T for complex **4**, < 1 T for complex **1**), the linear response becomes dominated by the fully reversible van Vleck susceptibility, which arises from the field-induced modulation of the spin wave functions.

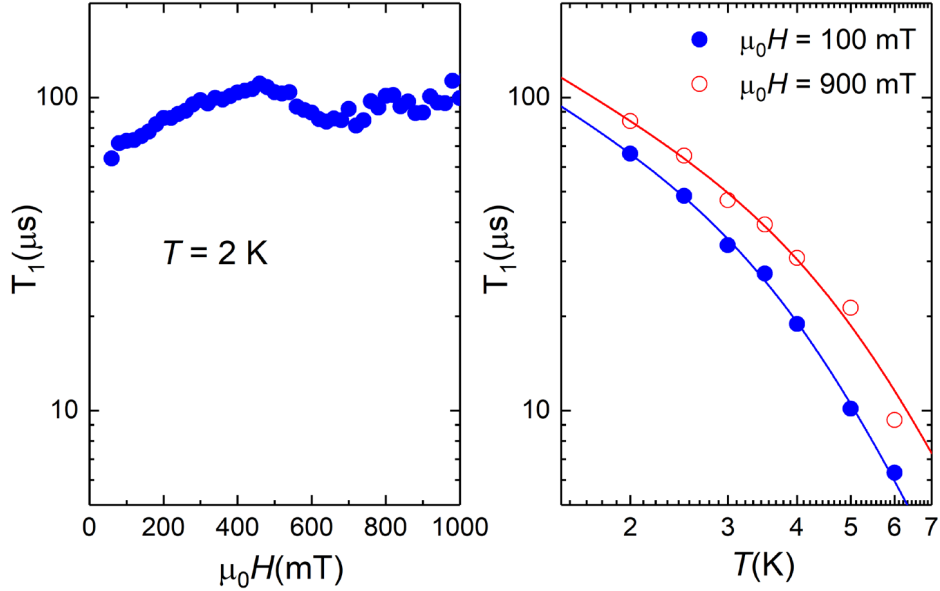

**Figure 11.** Spin-lattice relaxation time of **4** derived from frequency-dependent ac susceptibility experiments performed on a powdered sample as a function of magnetic field at $T$ = 2 K (left) and at two different magnetic fields as a function of temperature (right). The solid lines are fits that include direct and Raman relaxation processes (cf Eq. (5)).

In the case of complex **4**, we have been able to extract $T_1$ for temperatures below approximately 6 K and magnetic fields above 0.1 T. We have measured both a single crystal and a powder obtained by crushing it and embedding it in apiezon N grease, in order to rule out effects associated with a poor thermal contact between the sample and the sample holder (phonon bottleneck effect[47]). Representative $T_1$ data are shown in Fig. 11 (for the powder) and Fig. S21 (for the single crystal).

The results show that $T_1$ approaches 100 μs at $T$ = 2 K. The temperature dependence is compatible with relaxation driven by a combination of Raman and direct processes,[48,49] and can be described by the expression

$$\frac{1}{T_1} = A_{dir}T + A_{Raman}T^4 \quad (5)$$

The exponent of the Raman term in Eq. (5) has been fixed to the one that corresponds to non Kramers ions when optical phonons play a role,[49] as it is likely the case for molecular spins.[50] The weak dependence on magnetic field also agrees with a dominant Raman relaxation process.[48] However, the direct process might take over at sufficiently strong magnetic fields (when the gap between the ground and first excited levels becomes large enough), as can be seen in the results measured on the single crystal above 0.7-0.8 T (Fig. S21). This effect could also account for the faster relaxation of complex **1** and underlines, once more, the importance of tuning the quantum tunnelling splitting.

For complex **1**, a complementary estimate of $T_1$ has been obtained from high-frequency EPR experiments (Fig. S22). At $T$ = 2 K, the EPR spectrum shows saturation effects associated with a relatively slow spin-lattice relaxation (Fig. S23). From the power dependence of the signal intensity the spin-lattice relaxation time $T_1$ can be estimated to be in the order of tens of microseconds, thus in agreement with ac susceptibility results that point to $T_1 < 100$ μs (see SI for further details).

Let us compare now the spin-lattice relaxation times with typical quantum operation time scales. At the clock transition, the Rabi frequency for quantum oscillations between the tunnel split states $\Omega_R = g\mu_B \langle 0|\vec{b}\vec{S}|1\rangle/\hbar \approx 2g\mu_B b_z S/\hbar$, where $\vec{b}$ is the resonant microwave field that must have nonzero component along the anisotropy axis $z$ (otherwise, the transition is forbidden). For a typical $b_z$ = 1 mT, this results in qubit operation frequencies of the order of GHz (operation times of order $10^{-3}$ μs). Therefore, we can safely conclude that the spin-lattice relaxation will not limit the application of these molecules as spin qubits. The relevant time scale at low temperatures then becomes $T_2$, which will be maximized at the clock transition thanks to its insensitiveness to magnetic field fluctuations.[23-26]

## Conclusions

The results discussed in the previous sections neatly show the existence of a sizeable quantum tunnel splitting $\Delta$ in mononuclear Ni(II) complexes. Heat capacity experiments enable a direct detection of this quantum gap and monitoring how it evolves with magnetic field, following the quadratic dependence that characterizes a spin clock transition. The same technique evidences also that the molecular spin qubits effectively decouple from each other at the clock transition if the gap is larger than the typical energy scale of spin-spin interactions. The latter effect is of intrinsic interest for fundamental Physics, and can be understood within the framework of the quantum Ising model in a transverse magnetic field.[51] This paradigmatic model describes a quantum phase transition. For $\Delta > Z|J|S^2/2$, the spin lattice enters a quantum paramagnetic state and stays there, i.e. no transition to long-range order occurs, down to $T = 0$. A realization of such quantum phase transition has been observed for crystals of Fe$_8$ molecular clusters.[52] However, while in this case the transition was driven by an external magnetic field applied perpendicular to the magnetic anisotropy axis, the spins of complex **1** are already in the quantum regime even at zero field on account of its large $\Delta$. These molecules provide then model candidates, with parameters tuneable by chemical means, to investigate the dynamics of pure two-level systems[53] with a well-defined quantum ground state, of which very few pure realizations exist in Nature.

The ability to tune the quantum tunnelling gap via the chemical design of the molecular structure is also of relevance for the development of robust building blocks for scalable quantum technologies. Two obvious reasons, already mentioned, are the possibility to select the most appropriate frequency to optimally interface with the control electronics, even at zero field, and the ability to initialize these qubits by simply cooling them down to very low temperatures. However, there is a more subtle potential advantage that deserves to be mentioned. One of the characteristic, and most appealing, aspects of molecular based spin qubits is the ability to scale up quantum resources within each molecule, e.g. by introducing several spin sites each acting as a qubit. This strategy has given rise to model systems, able to implement basic quantum gates[7,11-13] or even quantum error correction algorithms[15] at the molecular scale. Decoupling the qubits from each other contributes to enhance the spin coherence. However, it also has a clear downside: a set of $N$ identical uncoupled qubits does not work as an $N$-qubit processor, as it lacks the possibility of individually addressing each qubit and of performing two-qubit gates. The ability to tune $\Delta$ provides both the possibility of making each qubit different (e.g. by having a different resonance frequency) and of activating the spin-spin interactions that are essential to implement conditional operations.[54] Molecular structures hosting several Ni(II) centres, which are chemically feasible, could therefore encode multiple addressable qubits while preserving resilience against decoherence. An illustrative example is given in Fig. S24 for the simplest situation of two coupled $S = 1$ spins.[55] Because of their chemical tuneability, the molecular complexes reported here are not just suitable qubits, but proper elements to build more complex quantum functionalities (a kind of "decoherence-free" quantum processor) at the molecular level.

## Conflicts of interest

There are no conflicts to declare.

## Acknowledgements

This work was supported by funds from the EU (COST Action 15128 MOLSPIN, QUANTERA project SUMO, FET-OPEN grant 862893 FATMOLS), the Spanish MICINN (grants RTI2018-096075-B-C21 and PCI2018-093116) and the Gobierno de Aragón (grant E09-17R-Q-MAD). TM thanks the IUF (Institut Universitaire de France) for financial support.

## Notes and references

1. M. N. Leuenberger and D. Loss, *Nature*, 2001, **410**, 789–793.
2. F. Troiani and M. Affronte, *Chem. Soc. Rev.*, 2011, **40**, 3119-3129.
3. G. Aromí, D. Aguilà, P. Gamez, F. Luis, and O. Roubeau, *Chem. Soc. Rev.*, 2012, **41**, 537-546.
4. E. Moreno-Pineda, C. Godfrin, F. Balestro, W. Wernsdorfer and M. Ruben, *Chem. Soc. Rev.*, 2018, **47**, 501-513.
5. A. Gaita-Ariño, F. Luis, S. Hill and E. Coronado, *Nat. Chem.*, 2019, **11**, 301–309.
6. M. Atzori and R. Sessoli, *J. Am. Chem. Soc.*, 2019, **141**, 11339-11352.
7. F. Luis *et al*, *Phys. Rev. Lett.*, 2011, **107**, 117203.
8. C. J. Wedge *et al*, *Phys. Rev. Lett.*, 2012, **108**, 107204.
9. M. J. Martínez-Pérez *et al*, *Phys. Rev. Lett.*, 2012, **108**, 247213.
10. D. Aguilà *et al*, *J. Am. Chem. Soc.*, 2014, **136**, 14215–14222.
11. A. Ardavan *et al*, *npj Quantum Inf.*, 2015, **1**, 15012.
12. A. Fernández *et al*, *Nat. Commun.*, 2016, **7**, 10240.
13. J. Ferrando-Soria *et al*, *Nat.Commun.*, 2016, **7**, 11377.
14. M. Atzori *et al*, *Chem. Sci.*, 2018, **9**, 6183–6192.
15. E. Macaluso *et al*, *Chem. Sci.*, 2020, **11**, 10337-10343.
16. A. Ardavan *et al*, *Phys. Rev. Lett., 2007,* **98**, 057201.
17. A. Morello, P. C. E. Stamp and I. S. Tupitsyn, 2006, *Phys. Rev. Lett.*, 2006, **97**, 207206.
18. S. Takahashi *et al.*, *Nature*, 2011, **476**, 76–79.
19. K. Bader *et al*, *Nat. Commun.*, 2014, **5**, 5304.
20. J. M. Zadrozny, J. Niklas, O. G. Poluetkov and D. E. Freedman, ACS Cent. Sci., 2015, **1**, 488–492.
21. M. Atzori *et al*, *J. Am. Chem. Soc.*, 2016, **138**, 2154–2157.
22. K. S. Pedersen, A.-M. Ariciu, Si. McAdams, H. Weihe∥, J. Bendix, F. Tuna and S. Piligkos, J. Am. Chem. Soc. 2016, **138**, 18, 5801–5804
23. G. Wolfowicz, A. M. Tyryshkin, R. E. George, H. Riemann, N. V. Abrosimov, P. Becker, H.-J. Pohl, M. L. W. Thewalt, S. A. Lyon and J. J. L. Morton, *Nature Nanotech.*, 2013, **8**, 561–564.
24. M. Shiddiq, D. Komijani, Y. Duan, A. Gaita-Ariño, E. Coronado and S. Hill, *Nature*, 2016, **531**, 348–351.
25. J. M. Zadrozny, A. T. Gallagher, T. D. Harris and D. E. Freedman, *J. Am. Chem. Soc.*, 2017, **139**, 7089–7094.
26. C. A. Collett, K.-I. Ellers, N. Russo, K. R. Kittilstved, G. A. Timco R. E. P. Winpenny and J. R. Friedman, *Magnetochemistry*, 2019, **5**, 4.
27. S. Giménez-Santamarina, S. Cardona-Serra, J. M. Clemente-Juan, A. Gaita-Ariño and E. Coronado, *Chem. Sci.*, 2020, **11**, 10718-10728.
28. A. Abragam and B. Bleaney, 1970, *Electron Paramagnetic Resonance of Transition Ions*.
29. R. Ruamps, R. Maurice, L. Batchelor, M. Boggio-Pasqua, R. Guillot, A. Barra, J. Liu, E. Bendeif, S. Pillet, S. Hill, T. Mallah and N. Guihéry, *J. Am. Chem. Soc.*, 2013, **135**, 3017–3026.
30. R. Ruamps, L. J. Batchelor, R. Guillot, G. Zakhia, A. Barra, W. Wernsdorfer, N. Guihéry and T. Mallah, *Chem. Sci.*, 2014, **5**, 3418–3424.
31. H. B. Callen, 1985, *Thermodynamics and an Introduction to Thermostatistics*.
32. F. Luis, F. L. Mettes, J. Tejada, D. Gatteschi and L. J. de Jongh, *Phys. Rev. Lett.*, 2000, **85**, 4377-4380.
33. M. Evangelisti, F. Luis, L. J. de Jongh and M. Affronte, *J. Mater. Chem.*, 2006, **16**, 2534-2549.
34. R. J. Schoelkopf and S. M. Girvin, *Nature*, 2008, **451**, 664–669.
35. M. D. Jenkins *et al*, *Dalton Trans.*, 2016, **45**, 16682-16693.
36. C. Bonizzoni, A. Ghirri, F. Santanni, M. Atzori, L. Sorace, R. Sessoli and M. Affronte, *npj Quantum Info.*, 2020, **6**, 68.

37  F. Neese, *WIREs Comput. Mol. Sci.*, 2012, **2**, 73–78.
38  R. Maurice, R. Bastardis, C. d. Graaf, N. Suaud, T. Mallah and N. Guihéry, *J. Chem. Theory Comput.*, 2009, **5**, 2977–2984.
39  C. Angeli, R. Cimiraglia, S. Evangelisti, T. Leininger and J. P. Malrieu, *J. Chem. Phys.*, 2001, **114**, 10252.
40  C. Angeli, R. Cimiraglia and J. P. Malrieu, *Chem. Phys. Lett.*, 2001, **350**, 297–305.
41  C. Angeli, R. Cimiraglia and J. P. Malrieu, *J. Chem. Phys.*, 2002, **117**, 913.
42  F. Neese, *J. Chem. Phys.*, 2005, **122**, 34107.
43  F. E. Mabbs and D. Collison, 1992, *Electron Paramagnetic Resonance of d Transition Metal Compounds*
44  N. Suaud, G. Rogez, J. Rebilly, M. Bouammali, N. Guihéry, A. L. Barra and T. Mallah, *Appl. Magn. Reson.*, 2020, **51**, 1215–1231.
45  K. S. Cole and R. H. Cole, *J. Chem. Phys.*, 1941, **9**, 341–352.
46  S. Gómez-Coca, A. Urtizberea, E. Cremades, P. J. Alonso, A. Camón, E. Ruiz and F. Luis, *Nat. Commun.*, 2014, **5**, 4300.
47  J. H. Van Vleck, *Phys. Rev.*, 1941, **59**, 724.
48  R. Orbach, *Proc. Roy. Soc.*, 1961, **A264**, 458.
49  K. N. Shrivastava, *Phys. Stat. Sol.*, 1983, **117**, 437.
50  A. Lunghi, F. Totti, S. Sanvito and R. Sessoli, *Chem. Sci.*, 2017, **8**, 6051-6059.
51  R. B. Stinchcombe, *J. Phys. C: Solid State Phys.*, 1973, **6**, 2459.
52  E. Burzurí, F. Luis, B. Barbara, R. Ballou, E. Ressouche, O. Montero, J. Campo and S. Maegawa, *Phys. Rev. Lett.*, 2011, **107**, 097203.
53  A. J. Leggett, S. Chakravarty, A. T. Dorsey, Matthew P. A. Fisher, Anupam Garg and W. Zwerger, *Rev. Mod. Phys.*, 1987, **59**, 1.
54  F. Luis, P. J. Alonso, O. Roubeau, V. Velasco, D. Zueco, D. Aguilà, J. I. Martínez, L. A. Barrios and G. Aromí, *Commun Chem.*, 2020, **3**, 176.
55  F. El-Khatib, B. Cahier, M. Lopez-Jorda, R. Guillot, E. Riviere, H. Hafez, Z. Saad, J. J. Girerd, N. Guihery and T. Mallah, *Inorg. Chem.*, 2017, **56**, 10655-10663.

# Electronic Supporting Information for

# Chemical tuning of spin clock transitions in molecular monomers based on nuclear spin-free Ni(II)

Marcos Rubín-Osanz,[a] François Lambert,[b] Feng Shao,[b] Eric Rivière,[b] Régis Guillot,[b] Nicolas Suaud,[c] Nathalie Guihéry,[c] David Zueco,[a] Anne-Laure Barra,[d] Talal Mallah,*[b] and Fernando Luis *[a]

## 1. Supplementary information

### 1.1. Synthesis

**N-méthyl N,N-bis{3-[(2-imidazolyl-méthyl)-amino]-propyl}- amine (2-imdipa)**: 1.09 g (0.0075 mol) of methy-di(3-aminopropyl)-amine are dissolvel in 20 ml of methanol. The solution is added dropwise under stirring to 1.5 g (0.0156 mol) of 2-carboxaldehyde imidazole dissolved in 45 ml of hot methanol. The obtained solution is refluxed (70 °C) for 15 minutes and left to cooldown to room temperature. 300 mg of activated Pd/C (10% Pd) are added very slowly to the solution. The mixture is loaded in a steel vessel and left to stirr at room temperature overnight under a $H_2$ pressure of 40 bars. The Pd/C is removed by filtration on cellite and thoroughly washed with methanol. The solution is evaporated under vacuum to give a yellow oil, with a yield of 98%. NMR 1H ($CDCl_3$): 1.55 ppm (m, 4H), 2,1 ppm (s, 3H), 2,3 ppm (t, 4H); 2,6 ppm (t, 4H) ; 3,85 ppm (s, 4H) ; 6-6,5 ppm (s large, 4H) ; 6,9 ppm (s, 4H).

**[Ni(2-imdipa)(NCS)][NCS]•(0.5MeOH) (4) :** 6.1 g of 2-imdipa (0.02 mol) are dissolved in 100 ml of methanol and added to a 100 ml methanolic solution containing 6.4 g (0.02 mol) of $[Ni(H_2O)_6]Cl_2$. The solution is stirred for 15 minutes at room temperature to which is added, under stirring, 200 ml of methanol containing 6.8 g (0.08 mol) of $NH_4NCS$. A change of color from blue-green to violet is observed. The solution is reduced to a volume of 150 ml under vacuum and 100 ml of water are added. The solution is then left to evaporate for three days. Blue-violet crystals form, they are collected by filtration washed within a minimum of distilled water and left to dry. The crystals are dissolved in hot methanol (150 ml) and filtered to eliminate a white precipitate. Yield 3 g (30%). Small crystals suitable for single crystal X-ray diffraction are obtained by ether diffusion into a methanolic solution of the complex. Large crystals (5x3x3 $mm^3$) are obtained by using the small crystals as seeds in a saturated methanolic solution of the complex under ether diffusion. Elemental analysis performed on microcystalline powder give for $NiC_{17.5}H_{29}N_{14}O_{0.5}S_2$ (Mw = 495.83) exp (calc.) %Ni: 11.41 (11.83), %C: 41.04 (42.35), %H:5.83 (5.85), %N: 25.78 (25.41), %S: 12.93 (13.01).

### 1.2. Crystallographic characterization of complex 4

X-ray diffraction data for compound **4** were collected on a Kappa X8 APPEX II Bruker diffractometer equipped with a graphite-monochromated Mo$_{K\alpha}$ radiation ($\lambda$ = 0.71073 Å). Crystal was mounted on a CryoLoop (Hampton Research) with Paratone-N (Hampton Research) as cryoprotectant and then flashfrozen in a nitrogen-gas stream at 100 K. The temperature of the crystal was maintained at the selected value by means of a 700 series Cryostream cooling device to within an accuracy of ±1 K. The data were corrected for Lorentz polarization and absorption effects. The structures were solved by direct methods using SHELXS-97 [1] and refined against $F^2$ by full-matrix least-squares techniques using SHELXL-2018 [2] with anisotropic displacement parameters for all non-hydrogen atoms. Hydrogen atoms were located on a difference Fourier map and introduced into the calculations as a riding model with isotropic thermal parameters. All calculations were performed by using the Crystal Structure crystallographic software package WINGX [3].

The crystal data collection and refinement parameters are given in Table S1.

The asymmetric unit of **4** consists of two independent molecules of [Ni(2-Imdipa)](NCS), two (NCS) and only one methanol.

CCDC 2000847 contains the supplementary crystallographic data for this paper. These data can be obtained free of charge from the Cambridge Crystallographic Data Centre via http://www.ccdc.cam.ac.uk/Community/Requestastructure.

### 1.3. Symmetry operations and magnetic field orientation for crystals of complex 4

The four molecule orientations in the unit cell of **4** are related by an inversion, a 2-fold screw axis along $\hat{b}$ and a glide plane generated by $\hat{a}$ and $\hat{c}$. Translations are irrelevant, so we can consider the 2-fold screw axis as a 2-fold rotation axis and the glide plane as a mirror plane. We ask now how a vector $\vec{x} = (x_1, x_2, x_3)$ transforms under these symmetry elements. If $x_1, x_2, x_3$ are the coordinates of $\vec{x}$ along $\hat{a}$, $\hat{b}$ and $\hat{a} \times \hat{b}$:

- Inversion:

$$i\vec{x} = \begin{pmatrix} -1 & 0 & 0 \\ 0 & -1 & 0 \\ 0 & 0 & -1 \end{pmatrix} \begin{pmatrix} x_1 \\ x_2 \\ x_3 \end{pmatrix} = -\vec{x}$$

- 2-fold rotation axis along $\hat{b}$:

$$C_2\vec{x} = \begin{pmatrix} -1 & 0 & 0 \\ 0 & 1 & 0 \\ 0 & 0 & -1 \end{pmatrix} \begin{pmatrix} x_1 \\ x_2 \\ x_3 \end{pmatrix} = \begin{pmatrix} -x_1 \\ x_2 \\ -x_3 \end{pmatrix}$$

- Mirror plane generated by $\hat{a}$ and $\hat{c}$ = mirror plane generated by $\hat{a}$ and $\hat{a} \times \hat{b}$:

$$\sigma\vec{x} = \begin{pmatrix} 1 & 0 & 0 \\ 0 & -1 & 0 \\ 0 & 0 & 1 \end{pmatrix} \begin{pmatrix} x_1 \\ x_2 \\ x_3 \end{pmatrix} = \begin{pmatrix} x_1 \\ -x_2 \\ x_3 \end{pmatrix}$$

The magnetic field $\vec{H}$ contributes to the Hamiltonian through the Zeeman term:

$$\mathcal{H}_{Zeeman} = -\vec{\mu} \cdot \vec{H}$$

Where $\vec{\mu}$ is the magnetic moment of the ion. The magnetic moments of the four molecule orientations are related through the symmetry operations described above. Then we have four different Zeeman terms:

- Inversion:

$$\mathcal{H}_{Zeeman} = -(i\vec{\mu}) \cdot \vec{H} = +\vec{\mu} \cdot \vec{H}$$

- 2-fold rotation axis along $\hat{b}$:

$$\mathcal{H}_{Zeeman} = -(C_2\vec{\mu}) \cdot \vec{H} = \mu_1 H_1 - \mu_2 H_2 + \mu_3 H_3$$

- Mirror plane generated by $\hat{a}$ and $\hat{c}$ = mirror plane generated by $\hat{a}$ and $\hat{a} \times \hat{b}$:

$$\mathcal{H}_{Zeeman} = -(\sigma\vec{\mu}) \cdot \vec{H} = -\mu_1 H_1 + \mu_2 H_2 - \mu_3 H_3$$

If only a global sign changes, the energy levels and states are the same. This happens if $\mu_2 = 0$ (the plane generated by *a* and *c*) or if $\mu_1 = \mu_3 = 0$ (the cell *b* axis). In these situations, the cosine directors of $\vec{H}$ in the molecular axes ($\vec{x} \cdot \vec{H}$, where $\vec{x}$ is any molecular axis) are also the same (but a global sign). From all these considerations we conclude that we only need to consider one molecule orientation if we measure along the cell b axis or the plane perpendicular to it.

## 2. Supplementary Tables

**Table S1**. Crystallographic data and structure refinement details of complex **4**.

| Compound | 4 |
|---|---|
| CCDC | 2000847 |
| Empirical Formula | 2($C_{16} H_{27} N_8$ Ni S), 2(C N S), C $H_4$ O |
| $M_r$ | 992.61 |
| Crystal size, $mm^3$ | 0.34  0.29  0.18 |
| Crystal system | monoclinic |
| Space group | $P\ 2_1/n$ |
| a, Å | 14.8018(4) |
| b, Å | 16.4349(4) |
| c, Å | 19.1348(5) |
| α, ° | 90 |
| β, ° | 102.4940(10) |
| γ, ° | 90 |
| Cell volume, $Å^3$ | 4544.6(2) |
| Z ; Z' | 4 ; 1 |
| T, K | 100(1) |
| Radiation type; wavelength Å | MoKα; 0.71073 |
| $F_{000}$ | 2088 |
| μ, $mm^{-1}$ | 1.064 |
| range, ° | 1.584 - 32.858 |
| Reflections collected | 145 386 |

| Reflections unique | 16 791 |
|---|---|
| $R_{int}$ | 0.0300 |
| GOF | 1.060 |
| Refl. obs. *I* > 2(*I*) | 14 103 |
| Parameters | 545 |
| wR$_2$ (all data) | 0.1084 |
| R value *I* > 2(*I*) | 0.0368 |
| Largest diff. peak and hole (e$^-$.Å$^{-3}$) | 1.943 ; -0.925 |

**Table S2:** Comparison of the bond lengths (in Å) and the angles (in °) of the two independent molecules:

| Complex | Ni(1) | Ni(2) |
|---|---|---|
| Ni - N1 | 2.0980(12) | 2.1136(12) |
| Ni - N2 | 2.1382(13) | 2.1457(13) |
| Ni - N3 | 2.1846(13) | 2.1760(13) |
| Ni - N4 | 2.1279(12) | 2.1311(12) |
| Ni - N5 | 2.0606(12) | 2.0803(12) |
| Ni - N6 | 2.0594(14) | 2.0605(14) |
| N1 - Ni - N2 | 79.66(5) | 78.38(5) |
| N1 - Ni - N3 | 174.40(5) | 174.71(5) |
| N1 - Ni - N4 | 91.69(5) | 93.11(5) |
| N1 - Ni - N5 | 90.59(5) | 89.07(5) |
| N1 - Ni - N6 | 87.36(5) | 88.14(5) |
| N2 - Ni - N3 | 96.92(5) | 97.20(5) |
| N2 - Ni - N4 | 169.20(5) | 170.55(5) |
| N2 - Ni - N5 | 92.88(5) | 95.23(5) |
| N2 - Ni - N6 | 89.64(6) | 88.94(5) |
| N3 - Ni - N4 | 92.19(5) | 91.50(5) |
| N3 - Ni - N5 | 94.03(5) | 94.26(5) |
| N3 - Ni - N6 | 88.18(5) | 88.88(5) |
| N4 - Ni - N5 | 80.67(5) | 80.38(5) |
| N4 - Ni - N6 | 96.45(5) | 94.94(5) |
| N5 - Ni -N6 | 176.42(5) | 174.42(5) |

**Table S3.** Contribution to $D$ and to $E$ of the 9 triplet and the 15 singlet excited states for complex **4** determined from *ab initio* calculations

| State | Contribution to $D$ | Contribution to $E$ |
|---|---|---|
| T0 | - | - |
| T1 | -34.809 | -0.868 |
| T2 | 16.805 | 5.163 |
| T3 | 14.84 | -4.708 |
| T4 | -0.014 | -0.018 |
| T5 | 0.008 | 0.005 |
| T6 | 0.002 | 0.005 |
| T7 | 0.002 | 0.001 |
| T8 | -0.001 | -0.001 |
| T9 | 0 | 0 |
| S0 | 0.001 | -0.003 |
| S1 | -0.002 | 0.002 |
| S2 | 13.188 | 0.063 |
| S3 | -5.882 | 3.68 |
| S4 | -6.769 | -3.817 |
| S5 | -0.002 | -0.005 |
| S6 | -0.021 | 0.001 |
| S7 | -0.001 | 0.025 |
| S8 | 0.009 | -0.002 |
| S9 | -0.021 | -0.022 |
| S10 | -0.37 | 0.309 |
| S11 | -0.328 | -0.41 |
| S12 | -0.286 | 0.196 |
| S13 | 0.962 | -0.023 |
| S14 | 0 | 0 |

## 3. Supplementary Figures

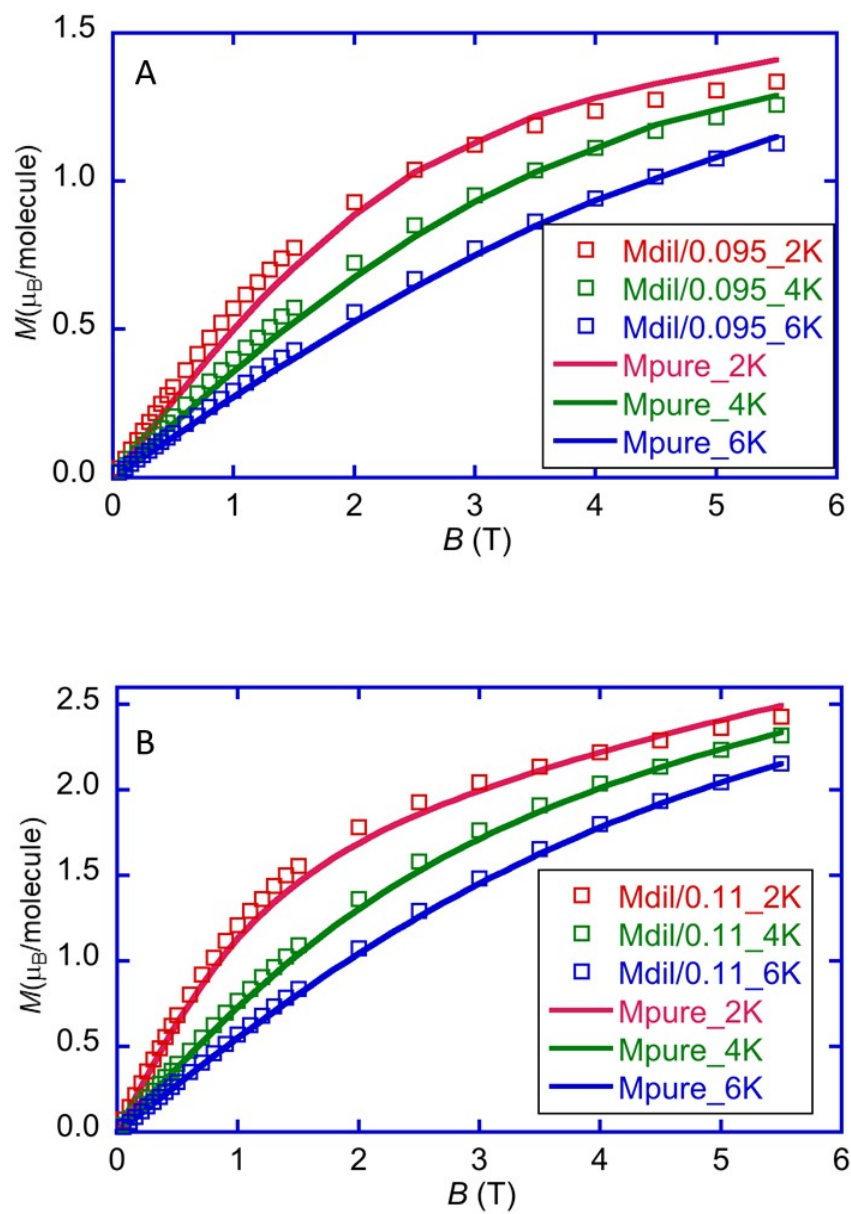

**Figure S1.** A: Magnetization isotherms measured on powder samples of **1** (lines) and of the magnetically diluted **1**$_{d9.5\%}$ sample (symbols) at the indicated temperatures. The scaling of the former data shows that the actual concentration is close to 9.5%. B: Magnetization isotherms measured on powder samples of **2** (lines) and of the magnetically diluted **2**$_{d11\%}$ sample (symbols) at the indicated temperatures. The scaling of the former data shows that the actual concentration is close to 11%.

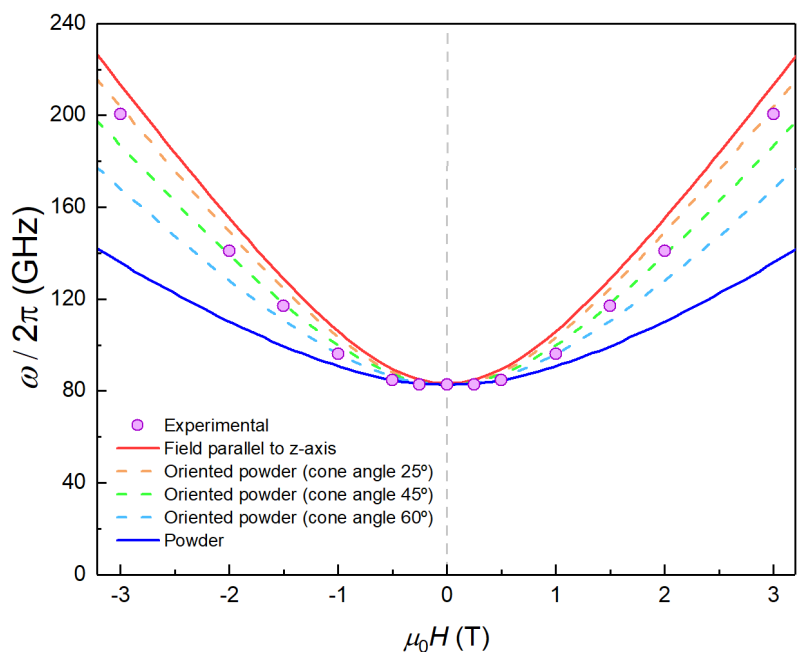

**Figure S2.** Experimental effective energy gap $\hbar\omega = k_B T_0/0.42$, with $T_0$ the field-dependent temperature of the heat capacity maximum measured on a powdered sample of **1** (Fig. 3 of the main text). Solid lines show the behaviour simulated for easy axes spread within a cone with varying aperture $\theta$. The limits $\theta = 0$ and $\theta = \pi/2$, shown in Fig. 3 of the main text, correspond to the anisotropy axes being perfectly aligned along the magnetic field and being randomly oriented, respectively.

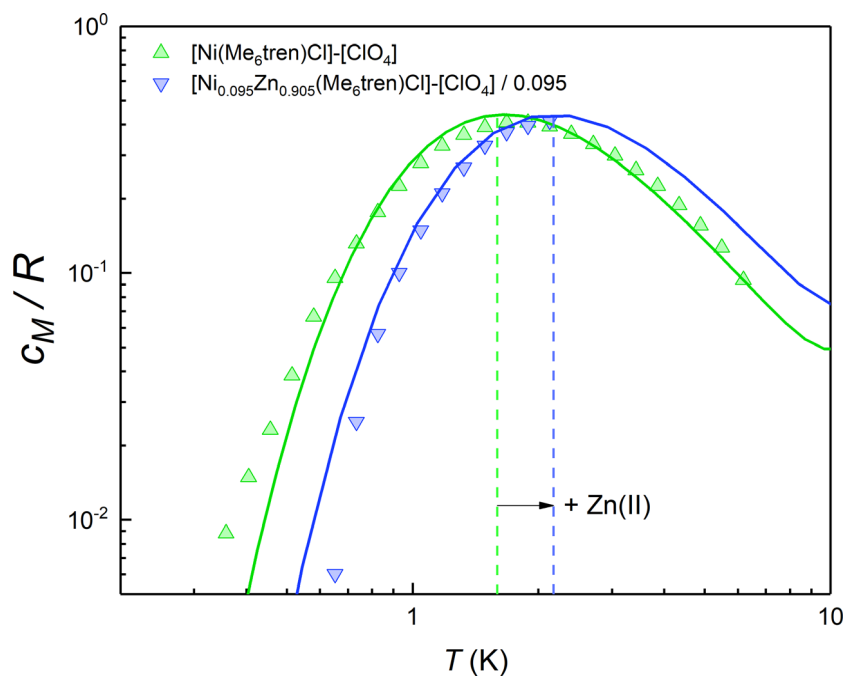

**Figure S3.** Specific heat of complexes **1** and **1**$_{d9.5\%}$ at $\mu_0 H = 0$. The lattice contribution has been subtracted, and the remaining magnetic contribution is normalised per mol of magnetic molecules. Despite the expected decrease in the strength of intermolecular magnetic interactions, these data show the same Schottky anomaly, even shifted to slightly higher temperatures. Therefore, the anomaly is due to the zero field splitting generated by the ligand field of each individual molecule.

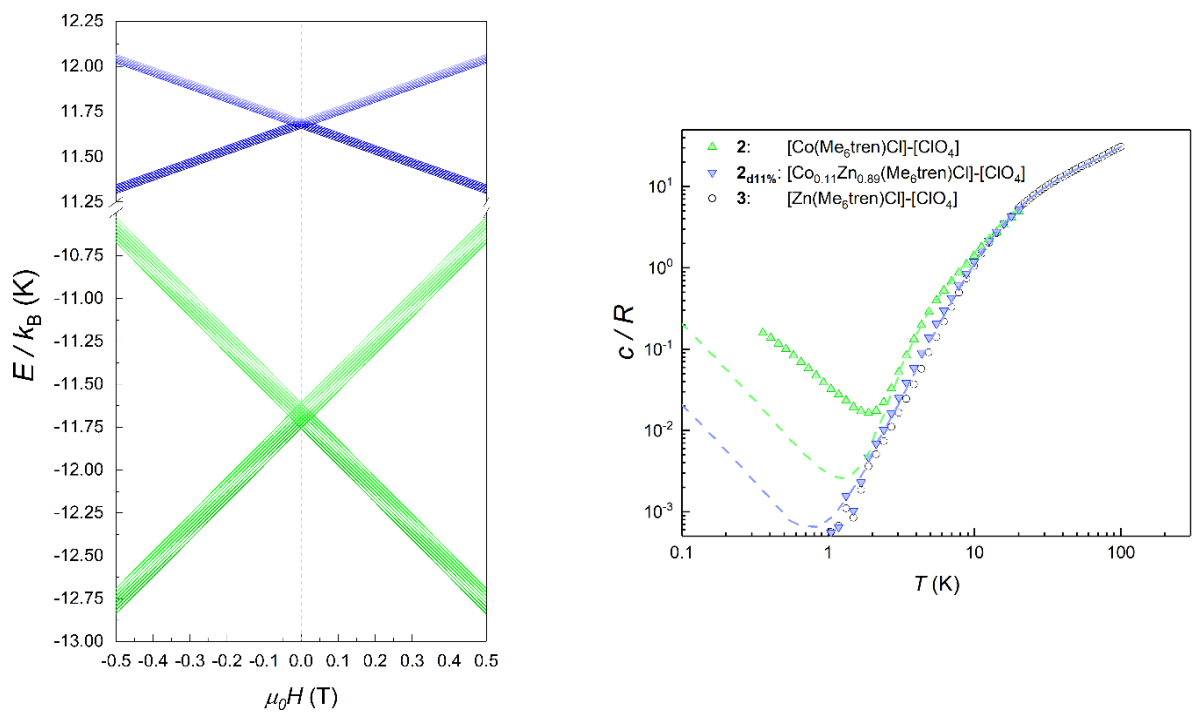

**Figure S4.** Left: Level structure of an isolated molecule of complex **2** including hyperfine interactions with a coupling constant $A_{hf}/k_B$ = 14 mK (see [4] and references therein) between the $S$ = 3/2 electronic spin and the $I$ = 7/2 nuclear spin. Right: The hyperfine level splitting gives rise to an additional contribution to the specific heat, which is shown by the dashed lines and compared to the experimental data obtained for the pure sample **2** and the magnetically diluted sample **2**$_{11\%}$. This comparison confirms that the hyperfine contribution cannot account for the specific heat of **2** below 1 K.

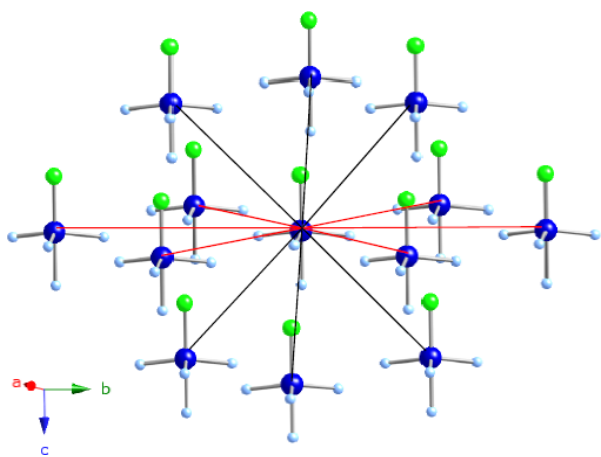

**Figure S5.** Location of the 12 nearest neighbour molecules of each [Co(Me$_6$tren)Cl]-(ClO$_4$) molecule in its crystal lattice, taken from reference [4]. This lattice has been used for the Monte Carlo calculations of spin-spin interactions, whose results are shown in Fig. 4 of the main text and in Fig. S4, and for the quantum "toy model" calculations shown in Fig. S6 below.

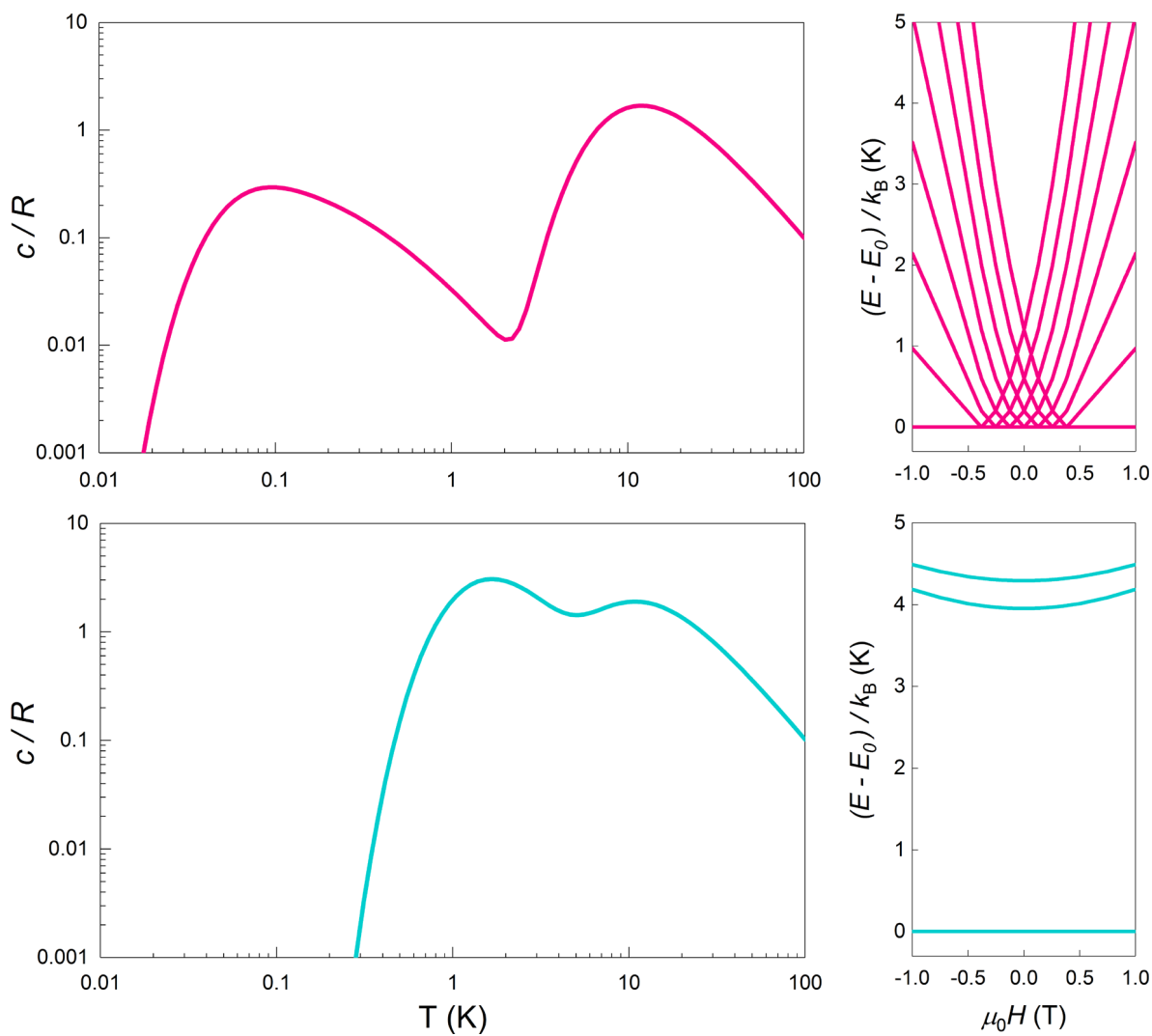

**Figure S6.** Results of quantum calculations performed on a set of 6+1 $S = 1$ spins located in the lattice shown in Fig. S5 and that include the central molecule plus its six nearest neighbours in the same crystal plane. The results have been obtained by numerical diagonalization of the multiple spin Hamiltonian, which includes the single molecule anisotropy (as given by Eq. (1) in the main text) and the couplings between different molecular spins (as given by Eq. (4) in the main text with $J = -0.035$ cm$^{-1}$ as determined for complex **2**). Panels at the top show the specific heat (left) and the field-dependent energy level structure (right, with energies referenced to the ground level) for the case of a vanishingly small quantum gap $\Delta$. The bottom panels show the calculations for $\Delta = 2.9$ cm$^{-1}$, as found for complex **1**. The presence of this gap lifts up in energy excitations due to spin-spin interactions and "clears up" their contribution to $c/R$ at very low temperature, just as it is observed in the experimental data measured on **1** (Fig. 2 of the main text).

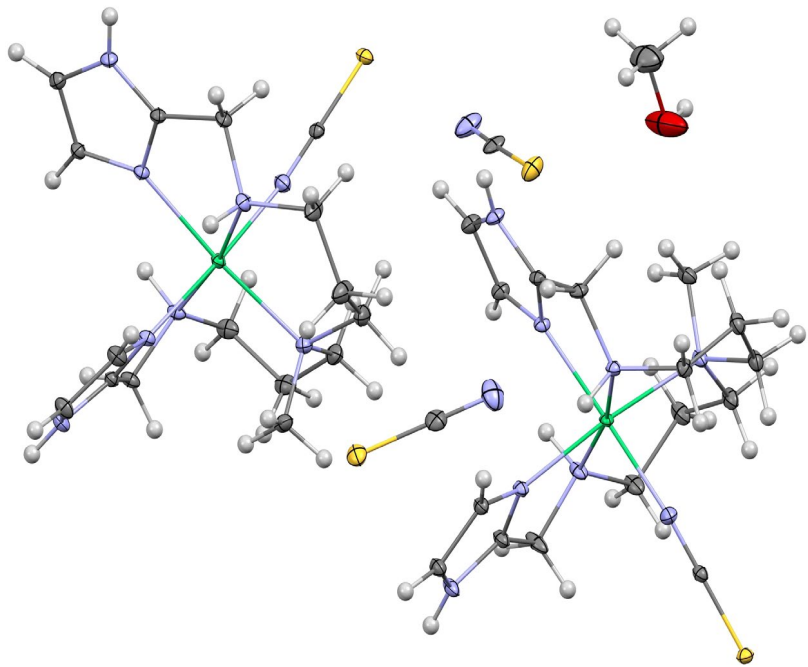

**Figure S7.** An ORTEP drawing of compound **4**. Thermal ellipsoids are shown at the 30% level.

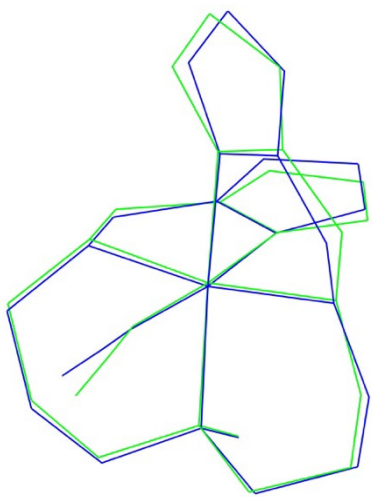

**Figure S8.** Schematic view of a superposition of the two crystallographically independent molecules

**Figure S9.** ORTEP drawing of complex **4** with atom numbering, H atoms are omitted for clarity

**Figure S10.** Thermal variation of $\chi T$ (○) experimental data and (—) best fit (see main text).

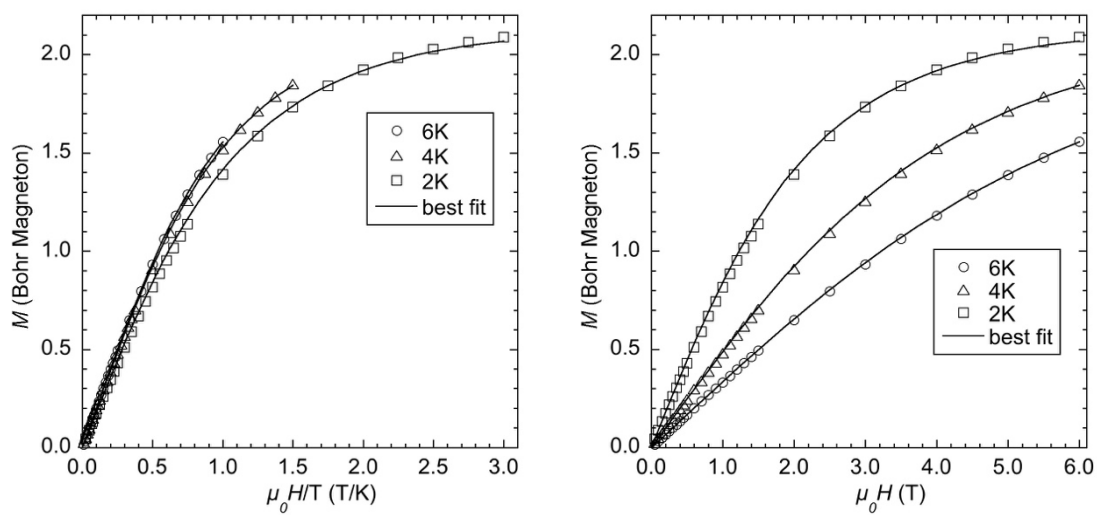

**Figure S11.** Magnetization (*M*) isotherms of a powder sample of **4** measured at *T* = 6 (□), 4 (△) and 2(○) K *versus* $\mu_0H/T$ (left) and *versus* $\mu_0H$ (right), (—) best fit (see main text).

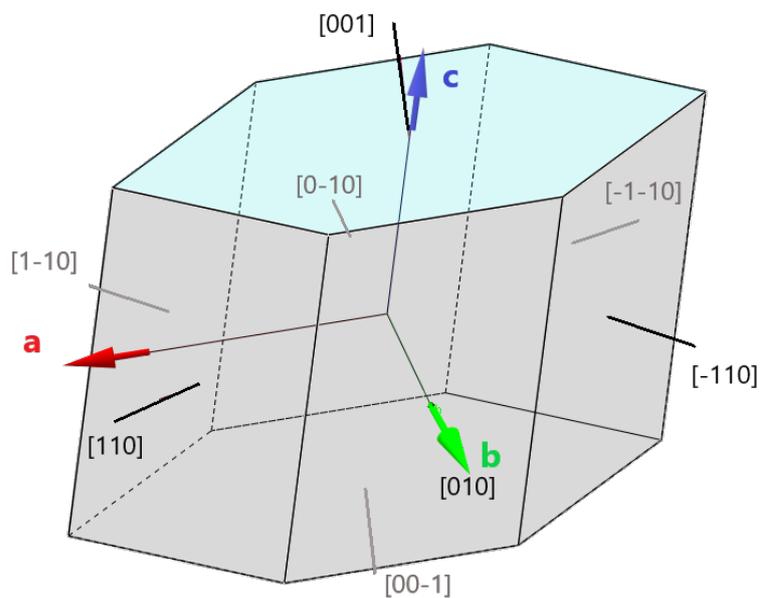

**Figure S12.** Schematic view of a single crystal of **4** with the crystallographic planes (001), (00-1), (010), (0-10), (110), (-1-10), (-110) and (1-10), which were modelled computationally using the crystal morphology editor/viewer (KrystalShaper) software (version 1.5.0 for Windows [5]) and the lattice parameters of **4**.

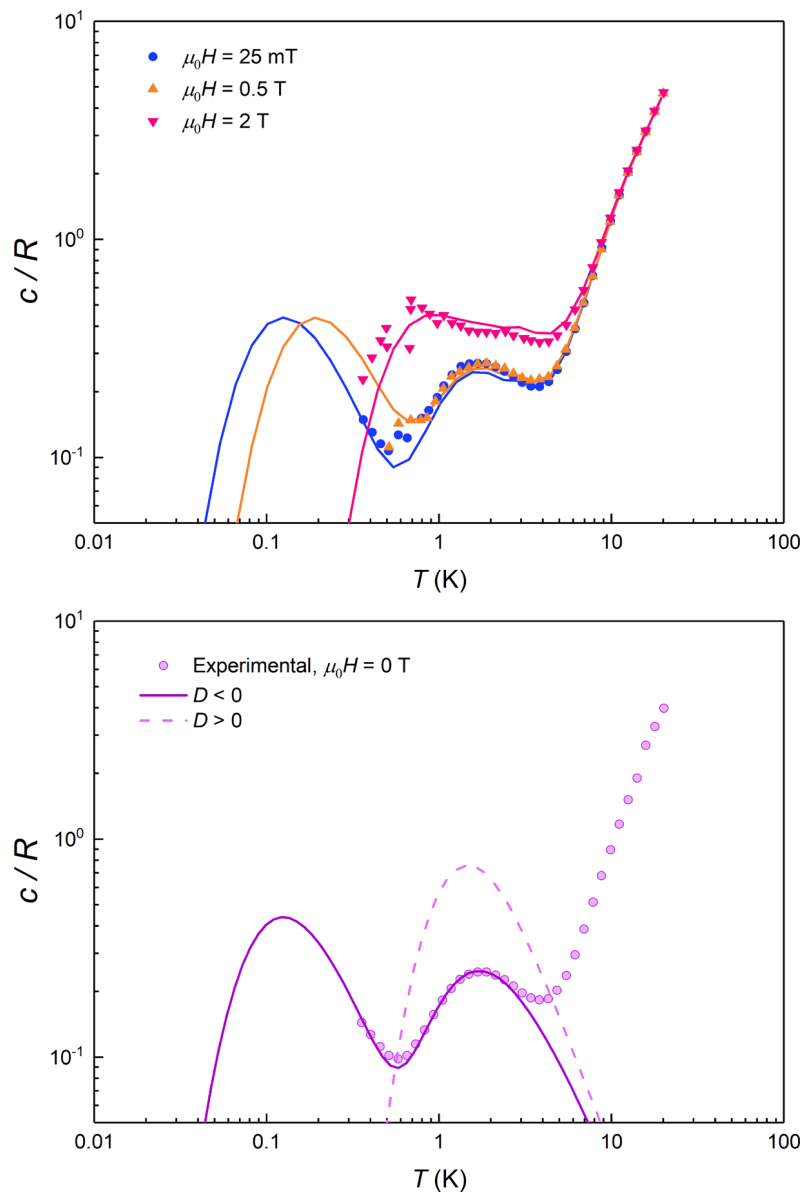

**Figure S13.** Top: Specific heat of a single crystal of **4** measured for magnetic fields 0.025 T (blue), 0.5 T (orange) and 2 T (pink) applied along the cell *b* axis. Solid lines show simulations performed for a magnetic field parallel to the magnetic *y* axis with the magnetic anisotropy parameters given in the main text. The agreement between experiment and theory shows that the molecular *y* axis points along the cell *b* axis. Bottom: Comparison between the zero-field specific heat data with simulations performed for positive and negative *D*, as indicated. The results unequivocally show that the magnetic anisotropy in **4** has a uniaxial character (*D* < 0).

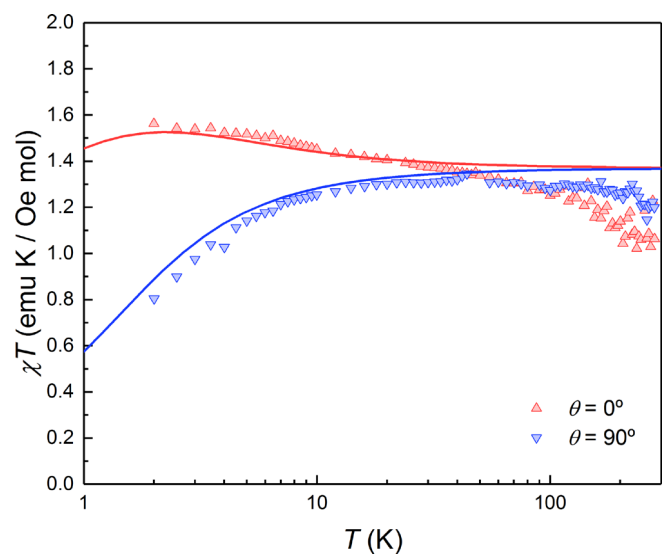

**Figure S14.** Magnetic susceptibility of a single crystal of **4** measured with the magnetic field applied along the orientations that give rise to a maximum (red) and a minimum (blue) magnetization in Figure 9 of the main text. Solid lines show simulations for the magnetic field applied along the *a* (red) and the *b* (blue) crystal axes, with the magnetic anisotropy parameters given in the main text.

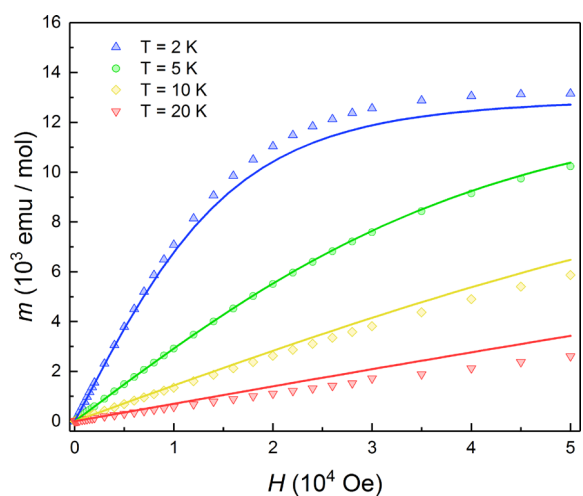

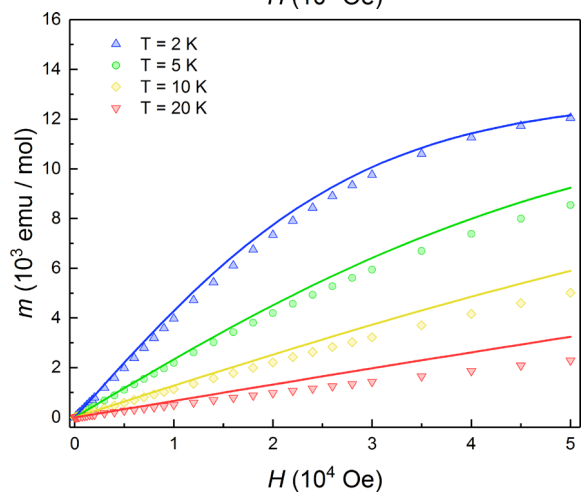

**Figure S15.** Magnetisation and magnetic susceptibility of a single crystal of **4** measured with the magnetic field applied along the maximum (left) and minimum (right) magnetisation orientations from Figure 9. Solid lines show the simulation of the magnetisation for the magnetic field applied along the cell *a* axis (left) and the cell *b* axis (right).

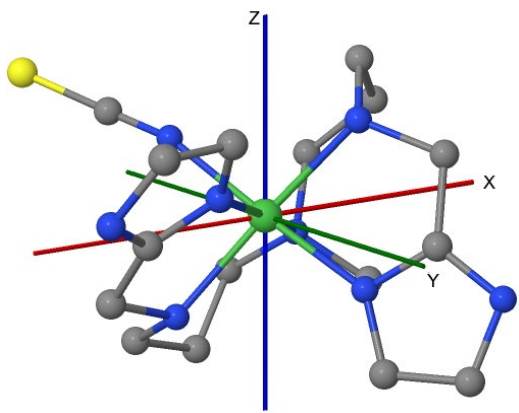

**Figure S16.** Orientation of the *D* tensor axes obtained from *ab initio* calculations, with *Z* (blue) the easy magnetization axis, *Y* (green) the intermediate magnetization axis and *X* (red) the hard magnetization axis.

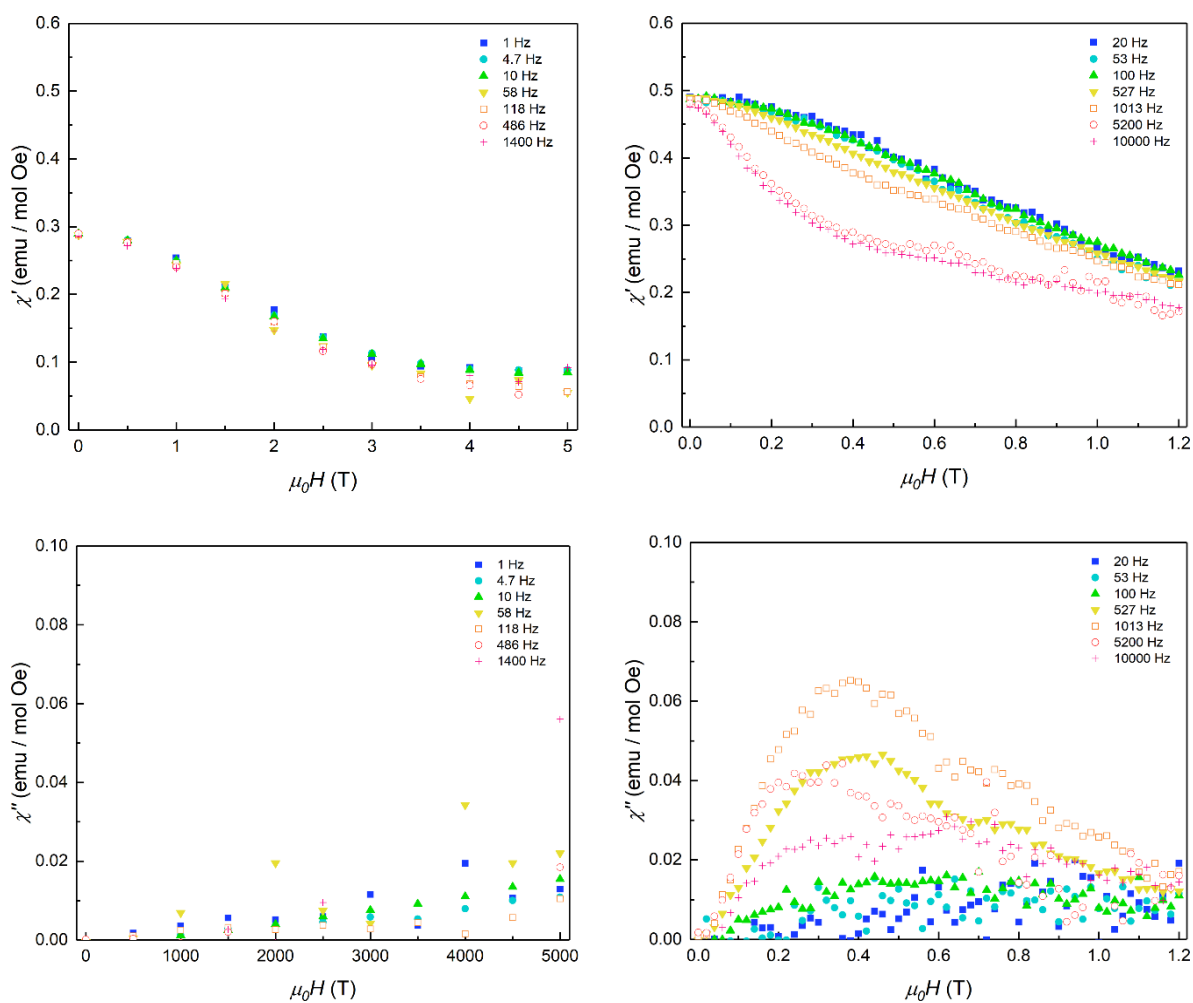

**Figure S17.** Ac susceptibility of powdered samples of **1** (left) and **4** (right) measured at T = 2 K as a function of magnetic field and for different frequencies $\omega/2\pi$ of the ac oscillating magnetic field. The top panels show the real component $\chi'$ and the bottom panels show the out-of-phase component $\chi''$. The dependence of $\chi'$ on $\omega$ and the onset of a non-zero $\chi''$ signal the existence of magnetic relaxation processes that 1) modify the magnetic response and 2) have relaxation time scales comparable to $1/\omega$. The results show that the irreversible susceptibility tends to vanish near zero field and that the spin relaxation is faster for complex **1**, which has a larger tunnelling splitting.

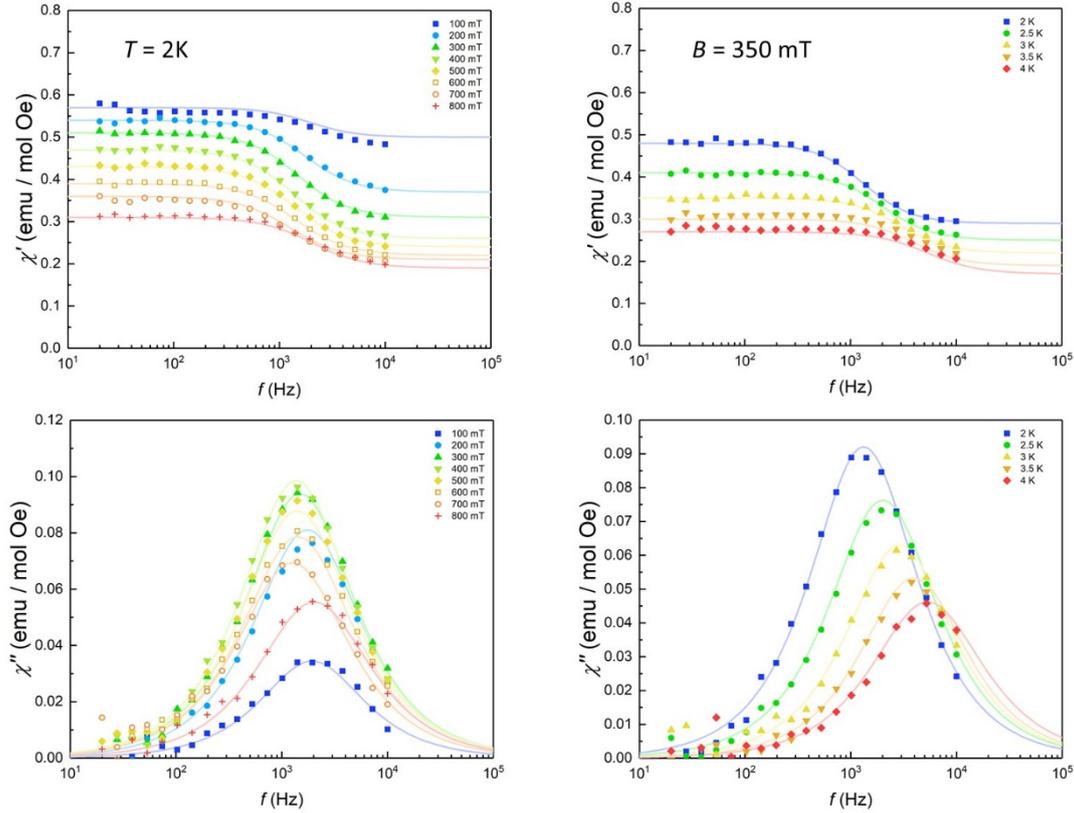

**Figure S18.** Frequency-dependent ac susceptibility of a single crystal of **4** measured at 2 K and different magnetic fields (left) and at $B$ = 350 mT and different temperatures (right). The dc and ac magnetic fields were applied along the crystallographic axis. Top and bottom panels show, respectively, the in-phase $\chi'$ and out-of-phase $\chi''$ susceptibility components. The symbols are experimental data and the lines are least-square fits performed with Cole-Cole functions [6]

$$\chi'(\omega) = \chi_S + (\chi_T - \chi_S)\frac{1 + (\omega\tau)^\beta \cos\left(\frac{\pi\beta}{2}\right)}{1 + 2(\omega\tau)^\beta \cos\left(\frac{\pi\beta}{2}\right) + (\omega\tau)^{2\beta}}$$

$$\chi''(\omega) = (\chi_T - \chi_S)\frac{(\omega\tau)^\beta \sin\left(\frac{\pi\beta}{2}\right)}{1 + 2(\omega\tau)^\beta \cos\left(\frac{\pi\beta}{2}\right) + (\omega\tau)^{2\beta}}$$

where $\omega$ is the angular frequency, $\chi_T$ the isothermal susceptibility, $\chi_S$ the adiabatic susceptibility $\tau$ the average spin relaxation time and $\beta$ describes a limited distribution of relaxation times (0.95-1 range).

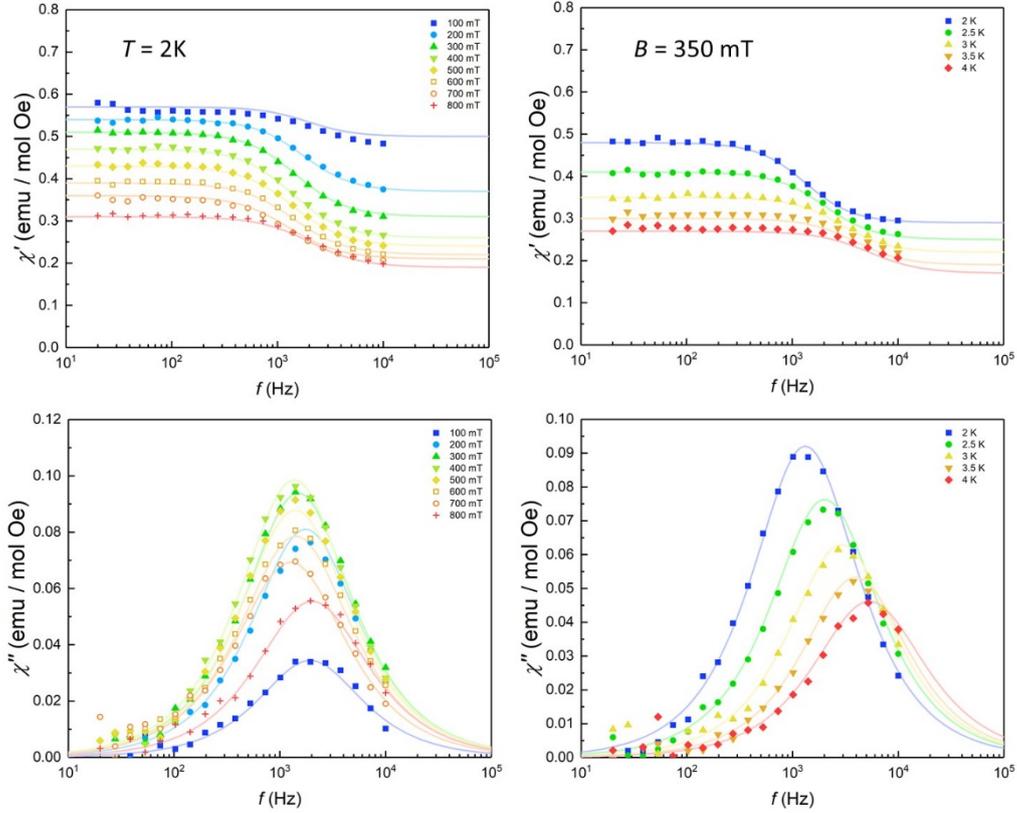

**Figure S19.** Frequency-dependent ac susceptibility of a powdered sample of **4** obtained by crushing the single crystal used in the measurements of Fig. S18 and mixing it in apiezon N grease. The measurements were then performed under identical conditions to those used for the single crystal. Top and bottom panels show, respectively, the in-phase $\chi'$ and out-of-phase $\chi''$ susceptibility components. The symbols are experimental data and the lines are least-square fits performed with Cole-Cole functions [6]

$$\chi'(\omega) = \chi_S + (\chi_T - \chi_S)\frac{1 + (\omega\tau)^\beta \cos\left(\frac{\pi\beta}{2}\right)}{1 + 2(\omega\tau)^\beta \cos\left(\frac{\pi\beta}{2}\right) + (\omega\tau)^{2\beta}}$$

$$\chi''(\omega) = (\chi_T - \chi_S)\frac{(\omega\tau)^\beta \sin\left(\frac{\pi\beta}{2}\right)}{1 + 2(\omega\tau)^\beta \cos\left(\frac{\pi\beta}{2}\right) + (\omega\tau)^{2\beta}}$$

where $\omega$ is the angular frequency, $\chi_T$ the isothermal susceptibility, $\chi_S$ the adiabatic susceptibility $\tau$ the average spin relaxation time and $\beta$ describes a limited distribution of relaxation times (0.95-1 range).

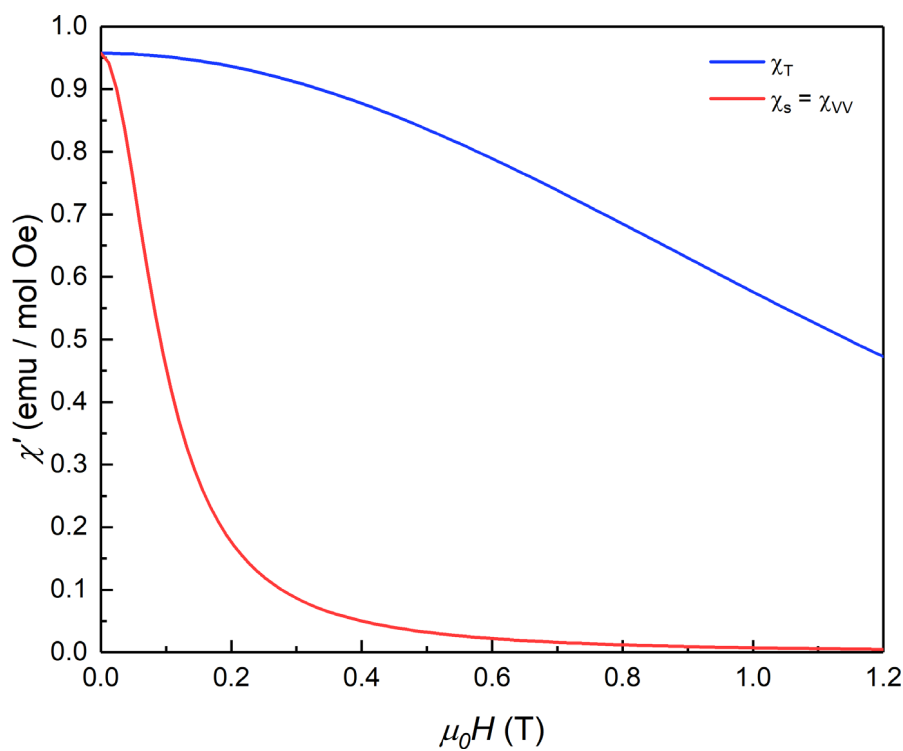

**Figure S20.** Magnetic field dependence of the equilibrium susceptibility $\chi_T$ and the reversible van Vleck susceptibility $\chi_S$ of complex **1** calculated from the spin Hamiltonian (1) with the anisotropy parameters given in the main text. The difference between the two represents the irreversible susceptibility, which is associated with phonon-induced transitions between different spin levels. Near the spin-clock transition at zero field, the spin wave functions are such that irreversible jumps between levels do not introduce any changes to the linear magnetic response. The ac susceptibility is then fully reversible and does not depend on frequency, even for spin lattice relaxation times $T_1 > 1/\omega$.

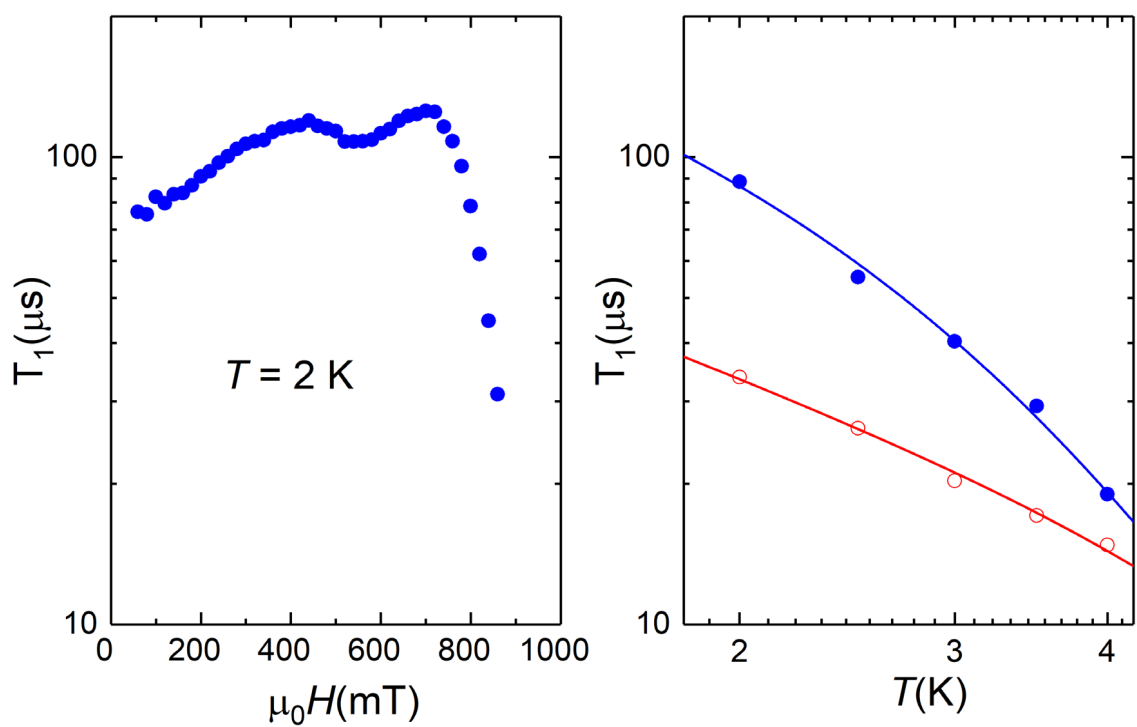

**Figure S21.** Spin-lattice relaxation time of **4** derived from frequency-dependent ac susceptibility experiments performed on a single crystal as a function of magnetic field at $T$ = 2 K (left) and at two different magnetic fields as a function of temperature (right). The magnetic field was aligned perpendicular to the *a* and *b* crystallographic axes, thus approximately at 52.6 degrees from the magnetic anisotropy axis *z* (see Fig. 9 in the main text). The solid lines are fits that include direct and Raman relaxation processes (cf Eq. (5) of the main text).

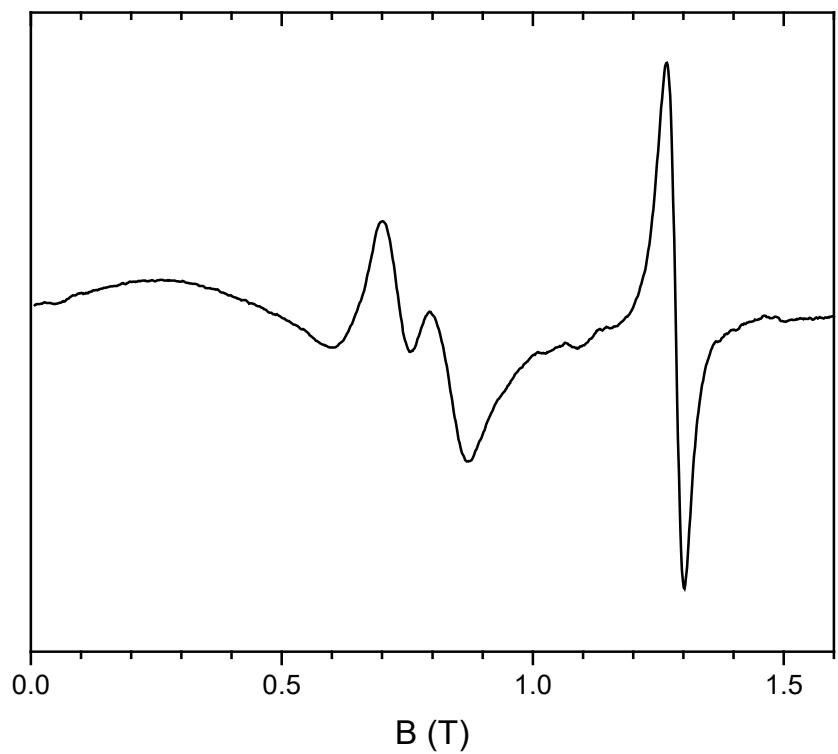

**Figure S22.** EPR spectrum of a large, but twinned, crystal of complex **4** measured at $T$ = 2 K, a frequency of 110.4 GHz and a microwave power equal to 9 mW. A broad and structured signal is observed in the 0.4-1T field range as well as a well-resolved line at 1.2 T with a gaussian shape and a linewidth of about 340 G at 2 K, and much larger at 5 K. For these frequency and magnetic fields, the transitions observed must involve the two lowest lying spin levels.

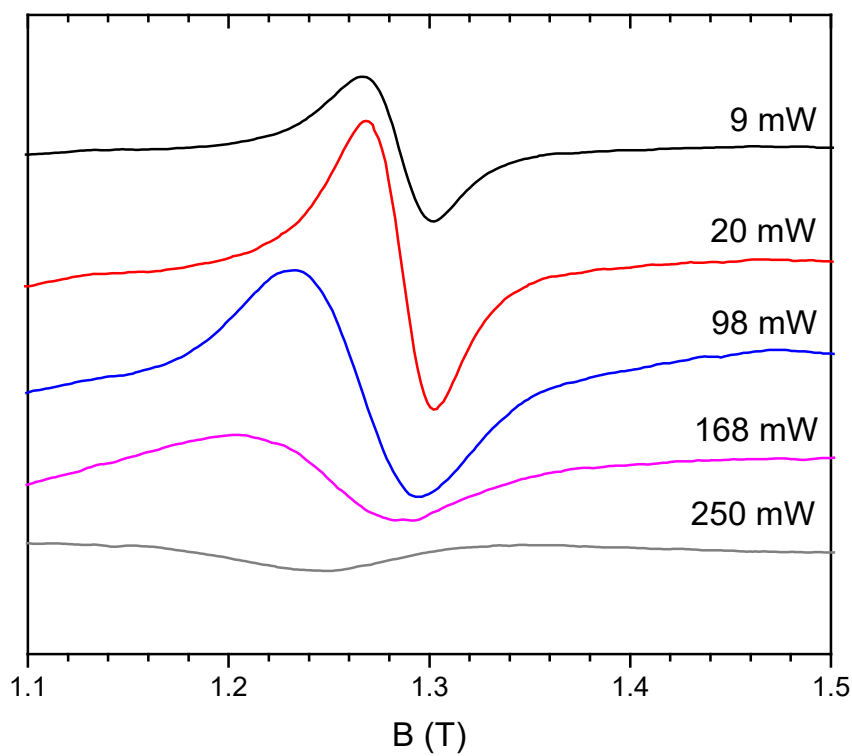

**Figure S23.** Change of the resonant signal observed near 1.2 T (cf Fig. S22) in the EPR spectrum of **1** as a function of the microwave power, for $T$ = 2 K and a frequency equal to 110.4 GHz. With the increase of the applied power (up to about 270 mW), the resonance line broadens and its height goes through a smooth maximum for power in the range 120-140 mW. These saturation effects show that $(\gamma b)^2 T_1 T_2$, where $b \approx$ 0.01-0.03 G is the microwave magnetic field amplitude, is no longer much smaller than 1. The resonance linewidth provides a lower bound for $T_2$ > 1-10 ns, although it is probably longer as the line shape shows the dominance of inhomogeneous broadening. The spin-lattice relaxation time $T_1$ can then be roughly estimated to be of the order of some tens of microseconds. More involved measurements are planned in order to better define $T_1$ (relying on longitudinal detection of the EPR signal) and eventually measure $T_2$.

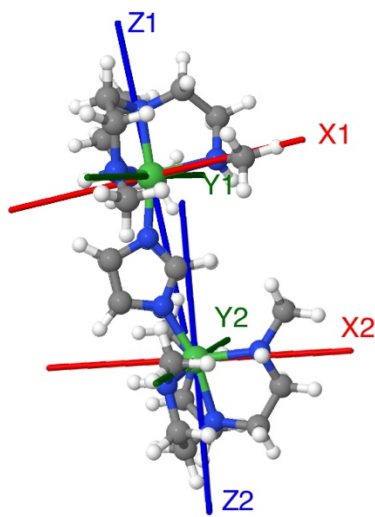

**Figure S24.** Schematic view of a Ni(II) binuclear complex with pentacoordinate geometry similar to that of **1** with the two Ni(II) having different spin Hamiltonian parameters ($D_1$, $E_1$ and $D_2$, $E_2$) and therefore different anisotropy tensor orientations [7].

## 4. References

[1] G. M. Sheldrick, SHELXS-97, Program for Crystal Structure Solution, University of Göttingen, Göttingen, Germany, **1997**.

[2] G. M. Sheldrick, *Acta Crystallogr. A* **2008**, *64*, 112-122.

[3] L. J. Farrugia, *J. Appl. Cryst*. **1999**, *32*, 837-838.

[4] R. Ruamps, L. J. Batchelor, R. Guillot, G. Zakhia, A. L. Barra, W. Wernsdorfer, N. Guihéry and T. Mallah, *Chem. Sci.* **2014**, *5*, 3418–3424.

[5] http://www.jcrystal.com/products/krystalshaper/

[6] K. S. Cole and R. H. Cole, *J. Chem. Phys.* **1941**, *9*, 341–352.

[7] F. El-Khatib, B. Cahier, M. Lopez-Jorda, R. Guillot, E. Riviere, H. Hafez, Z. Saad, J. J. Girerd, N. Guihery and T. Mallah, *Inorg. Chem.*, 2017, **56**, 10655-10663.